\documentclass[letterpaper,titlepage,11pt]{article}
\usepackage{hyperref}
\usepackage{amssymb,amsmath,amsfonts}
\usepackage{epsfig}

\def\clock{{\count0=\time
           \divide\count0 60
           \ifnum\count0<10 0\fi\the\count0
           \multiply\count0 -60 \advance\count0 \time
           :\ifnum\count0<10 0\fi \the\count0
         }}
\newcommand{\timestamp}{{\small\vbox{\hbox{\tt\jobname.tex}
\hbox{\the\day/\the\month/\the\year, \clock}}}}

\newcommand{\CB}{\mathcal{B}}
\newcommand{\CC}{\mathcal{C}}

\newcommand{\CH}{\mathcal{H}}

\newcommand{\CL}{\mathcal{L}}
\newcommand{\CM}{\mathcal{M}}
\newcommand{\CN}{\mathcal{N}}
\newcommand{\CO}{\mathcal{O}}
\newcommand{\CP}{\mathcal{P}}

\newcommand{\Z}{\mathbb{Z}}
\newcommand{\C}{\mathbb{C}}

\newcommand{\nn}{\nonumber}

\newcommand{\spa}{\ , \ \ }

\newcommand{\ds}{\displaystyle}

\newcommand{\matrto}[4]{\left( \begin{array}{cc} #1 & #2 \\
#3 & #4 \end{array} \right) }

\newcommand{\ads}{\mbox{AdS}}

\numberwithin{equation}{section}

\begin{document}

\begin{titlepage}

\rightline{\vbox{\small\hbox{\tt NORDITA-2009-77} }}
 \vskip 1.8 cm

\centerline{\LARGE \bf Full Lagrangian and Hamiltonian for quantum strings }
\vskip 0.2cm
\centerline{\LARGE \bf on $\ads_4\times \C P^3$ in a near plane wave limit} \vskip 1.5cm

\centerline{\large {\bf Davide Astolfi$\,^{1}$}, {\bf Valentina
Giangreco M. Puletti$\,^{2}$}, {\bf Gianluca Grignani$\,^{1}$},}
\vskip 0.2cm \centerline{\large {\bf Troels Harmark$\,^{2}$} and
{\bf Marta Orselli$\,^{3}$} }

\vskip 0.5cm

\begin{center}
\sl $^1$ Dipartimento di Fisica, Universit\`a di Perugia,\\
I.N.F.N. Sezione di Perugia,\\
Via Pascoli, I-06123 Perugia, Italy\\
\vskip 0.4cm
\sl $^2$ NORDITA\\
Roslagstullsbacken 23,
SE-106 91 Stockholm,
Sweden \vskip 0.4cm
\sl $^3$ The Niels Bohr Institute  \\
\sl  Blegdamsvej 17, DK-2100 Copenhagen \O , Denmark \\
\end{center}
\vskip 0.5cm

\centerline{\small\tt astolfi@pg.infn.it,
valentina@nordita.org, grignani@pg.infn.it,}
\centerline{\small\tt harmark@nordita.org, orselli@nbi.dk}

\vskip 1.5cm \centerline{\bf Abstract} \vskip 0.2cm \noindent We
find the full interacting Lagrangian and Hamiltonian for quantum
strings in a near plane wave limit of $\ads_4\times \C P^3$. The
leading curvature corrections give rise to cubic and quartic terms
in the Lagrangian and Hamiltonian that we compute in full. The
Lagrangian is found as the type IIA Green-Schwarz superstring in the
light-cone gauge employing a superspace construction with 32 grassmann-odd coordinates. The light-cone gauge for the fermions is
non-trivial since it should commute with the supersymmetry
condition. We provide a prescription to properly fix the $\kappa$-symmetry gauge condition to make it consistent with light-cone gauge.
We use fermionic field redefinitions to find a simpler
Lagrangian. To construct the Hamiltonian a Dirac procedure is needed in order to
properly keep into account the fermionic second class constraints.  We combine the field redefinition with a shift of the fermionic
phase space variables that reduces Dirac brackets to Poisson brackets. This
results in a completely well-defined and explicit expression for the
full interacting Hamiltonian up to and including terms quartic in
the number of fields.

\end{titlepage}

\small
\tableofcontents
\normalsize
\setcounter{page}{1}

\section{Introduction and summary}

For the last decade, the duality between four-dimensional $\CN=4$
superconformal Yang-Mills (SYM) theory and type IIB string theory on
$\ads_5\times S^5$ has been celebrated as the one example of an
exact duality between gauge theory and string
theory~\cite{Maldacena:1997re}. Only last year Aharony, Bergman,
Jafferis and Maldacena (ABJM), inspired by earlier work on superconformal
Chern-Simons theories~\cite{Schwarz:2004yj},
  proposed a new exact
duality between a Chern-Simons-matter gauge theory and M-theory
compactified on  $\ads_4\times S^7/\Z_k$~\cite{Aharony:2008ug}. In a
particular limit the gauge theory is dual to type IIA string
theory compactified on $\ads_4\times \C P^3$. In this region in the
parameter space, the new duality is between three-dimensional
$\CN=6$ superconformal Chern-Simons theory (ABJM theory) and type
IIA string theory on $\ads_4\times \C P^3$ which preserves 24 out
of the 32 supersymmetries. ABJM theory has $U(N)\times U(N)$ gauge
symmetry with Chern-Simons like kinetic terms at level $k$ and $-k$
and it is weakly coupled when the 't Hooft coupling $\lambda = N/k$
is small. Instead type IIA string theory on $\ads_4\times \C P^3$ is
a good description when $1 \ll \lambda \ll k^4$.

As for the  $\ads_5/\mbox{CFT}_4$ duality, there is some
evidence for integrability in the planar limit also for the $\ads_4 /\mbox{CFT}_3$
duality~\cite{Minahan:2008hf,Gaiotto:2008cg,Arutyunov:2008if,Stefanski:2008ik,Grignani:2008is,Gromov:2008qe,Astolfi:2008ji,Bak:2008cp,Sundin:2008vt,Kristjansen:2008ib,Zwiebel:2009vb,Minahan:2009te}. In particular an all-loop asymptotic Bethe ansatz has
been proposed~\cite{Gromov:2008qe,Ahn:2008aa}.  Recently a set of functional equations  in the form of a Y-system based on
the integrability of the superstring $\sigma$-model, which defines the anomalous dimensions of local single trace operators, has been formulated also for the $\ads_4 /\mbox{CFT}_3$ duality~\cite{Gromov:2009tv}.

One of the most important results in the study of integrability of
the $\ads_5/\mbox{CFT}_4$ correspondence was the calculation of the
complete set of first-order curvature corrections to the spectrum of
light-cone gauge string theory that arises in the expansion of
$\ads_5\times S^5$ about the plane-wave
\cite{Callan:2003xr,Callan:2004uv}, the so-called ``near BMN limit"
\cite{Berenstein:2002jq}. Among other results, this has produced the
first evidence of the famous ``three loop discrepancy", which was
then understood and solved by the inclusion of the dressing factor that
interpolates between weak and strong coupling, in the Bethe
equations that describe the spectrum of the gauge and the string
theory~\cite{Arutyunov:2004vx}.

Analogous calculations for the $\ads_4/\mbox{CFT}_3$ duality were
initiated in \cite{Astolfi:2008ji} (see also~\cite{Sundin:2008vt}) where the spectrum of two
bosonic oscillator states in the $SU(2)\times SU(2)$ sector was
computed and compared with the solutions of the proposed all loop
Bethe equations~\cite{Gromov:2008qe,Astolfi:2008ji}.
In order to
perform a complete analysis of the spectrum of string oscillators
around a pp-wave limit of $\ads_4\times \C P^3$ it is however
necessary to include all the bosonic and fermionic directions in the computation of the
interacting Lagrangian and Hamiltonian.

In this paper, using the $\ads_4\times \C P^3$ superspace construction for type IIA superstrings of refs.~\cite{Gomis:2008jt,Grassi:2009yj}, we provide the full interacting Lagrangian and
Hamiltonian for the type IIA Green-Schwarz superstring in the light
cone gauge in a near plane wave limit of $\ads_4 \times \C P^3$. The
near plane wave background is given by the leading pp-wave
background plus the $1/R$ and $1/R^2$ curvature corrections, $R$ being the radius of $\C P^3$. We find all the
terms in the Lagrangian and Hamiltonian which are quadratic, cubic
and quartic in the number of fields. The Penrose limit defining the
near plane wave background follows a null geodesic that moves along
one of the isometries of $\C P^3$~\cite{Grignani:2008is}. We can
thus use our Hamiltonian to find the leading finite-size corrections
to the spectrum of a string fluctuating around the null geodesic.

In the quantization of type IIA string theory on the  $\ads_4 \times \C P^3$ background there are many non trivial issues that should be carefully addressed and solved
in order to have a complete characterization of the spectrum.

\begin{itemize}
\item[(a)] The $\kappa$-symmetry gauge choice for the light-cone gauge on this background cannot be chosen as in $\ads_5\times S^5$ or in flat space. The less than maximal supersymmetry means that the light-cone quantization for the
fermions will be completely different than that for the $\ads_5\times S^5$ background since it should commute with the supersymmetry condition.
\item[(b)]
Fermionic field redefinitions are conveniently used to find a simpler
Lagrangian such that the
fermionic momenta have no $1/R^2$ corrections. This simplifies the phase space variables and makes it easier to construct the Hamiltonian.
\item[(c)] A Dirac procedure is needed in order to
properly keep into account the fermionic second class constraints. This is highly non-trivial in this case since there are both first and second order curvature corrections to the Hamiltonian that should be taken into account. The Dirac brackets can then be reduced to Poisson brackets by a suitable field
redefinition.
\item[(d)] We prove that the two-fermion terms in
the Lagrangian that arises from the superspace model, match those found using the general type IIA
Green-Schwarz Lagrangian of~ \cite{Cvetic:1999zs}.
\item[(e)]  There is strong evidence that the terms in the Hamiltonian which are quartic in the fields are not normal ordered. This is because there are divergences coming from the cubic terms in the Hamiltonian when used at the second order in perturbation theory for certain string states~\cite{Astolfi:2008ji, futurework} which can only be canceled by the inclusion of the appropriate normal ordering functions.
\end{itemize}

Let us comment on these points.
With respect to (a), the $\kappa$-symmetry gauge condition  $\Gamma^+\theta =0$ where $\theta$ is a 32-component spinor and $\Gamma^+ =\Gamma^0  + \Gamma^9$,  which is used for flat space and $\ads_5\times S^5$, cannot be used indiscriminately on $\ads_4\times\C P^3$.
The isometry group of $\C P^3$ is $SU(4)$, it preserves not only the $\C P^3$ metric, $g_{ab}$, but also the K\"ahler form, $J_{ab}$. Thus the vacuum is $SU(4)$ invariant. This symmetry of the ten dimensional theory is dual to the $R$-symmetry of the  $\mathcal{N}=6$ supersymmetry of the gauge theory.
To see how supersymmetry is realized one can study supersymmetry transformation laws~\cite{Nilsson:1984bj}. In the purely bosonic ground state of the 10-dimensional theory given by $\ads_4\times \C P^3$ the fermion fields are set to zero, thus the criterion for unbroken supersymmetry of the vacuum is that their supersymmetric variations should also vanish, $\delta\theta=0$. In order to see when this condition is realized, one can define a quantity out of the $SU(4)$ invariant K\"ahler form, which takes the following explicit form in terms of 32 dimensional gamma matrices
\begin{equation}
\label{defQ} J = \Gamma_{0123} \Gamma_{11} ( - \Gamma_{49} -
\Gamma_{56} + \Gamma_{78} ) = \Gamma_{5678} - \Gamma_{49}
(\Gamma_{56} - \Gamma_{78})
\end{equation}
where we choose for $ \C P^3$ the directions $4$ to $9$. One can
easily show that $J^2 = 2J + 3$ and hence that $J$ has 24
eigenvalues -1 and 8 eigenvalues 3.  It was shown already
in~\cite{Nilsson:1984bj} that in the vacuum defined by $\ads_4\times
\C P^3$, the condition $\delta\theta=0$ is realized only by the 24
eigenvalues $J=-1$. This is the reason why $\ads_4\times\C P^3$
preserves only 24-supersymmetries. The projector on to supersymmetric states is then simply $P=(3-J)/4$. Now, the light-cone gauge condition  $\Gamma^+\theta=0$ does not commute with $J$, and thus with the projector $P$, and there is no choice of path along  $\C P^3$ for which this could happen since the $\Gamma^a$ with $a=4,5,6,7,8,9$  do not commute with $P$.
Thus on $\ads_4\times\C P^3$ the standard light-cone gauge
condition is not always consistent with supersymmetry.

We shall show in this paper how to properly fix the $\kappa$-symmetry gauge condition on this
background in such a way that it is consistent with our choice of light-cone gauge.
This will be done for string states in a Penrose limit defining the
near plane wave background that follows a null geodesic that moves
along one of the isometries of $\C P^3$ \cite{Grignani:2008is}.
In particular even to derive the
pp-wave spectrum one should in principle use the appropriate
light-cone gauge condition. All the papers in the literature that
derived the pp-wave spectrum on this
background~\cite{Sugiyama:2002tf,Hyun:2002wu,Nishioka:2008gz,Gaiotto:2008cg}
used the standard $\Gamma^+\theta=0$ condition that, even if provides the correct spectrum for the pp-wave, is, in principle, inconsistent with $\kappa$-symmetry. Whereas for the
pp-wave spectrum this has proven not to be a real issue, a correct,
supersymmetry preserving, gauge fixing becomes crucial in deriving
the curvature corrections to the spectrum.

 As stated in point (b) it is convenient to first perform field redefinitions on the Lagrangian such that the fermionic momenta have no $1/R^2$ corrections.
This  has the advantage that one needs only to take into account the $1/R$ corrections to the fermionic
momenta when changing variables to fermionic phase space variables in the Hamiltonian and when performing the Dirac procedure.

With respect to (c), the set of constraints that arise from the definitions of the fermionic momenta for the Green-Schwarz type IIA superstring on the $\ads_4\times\C P^3$ background are second-class. This means that to make a
consistent quantization the quantum anticommutator of two fermionic fields should
be identified with their Dirac bracket (which depends on the Poisson bracket algebra of the
constraints) rather than with their Poisson bracket.
For type IIA superstrings on $\ads_4\times\C P^3$, at variance with what happens for type IIB superstrings on the $\ads_5 \times S^5$ background, the canonical commutation relations have a complicated structure due to the fact that the Dirac brackets receive both $1/R$ and $1/R^2$ corrections. This makes canonical quantization of the Hamiltonian a much harder problem.
Fortunately, one can circumvent this by making a field redefinition of the fermionic phase space variables which thus changes the Hamiltonian. We find in this paper a particularly elegant way to make this field redefinition, which is both first- and second-order in the curvature correction, by combining it with the initial field redefinition that one should perform to write out the Hamiltonian in fermionic phase space variables in going from the Lagrangian to the Hamiltonian. Moreover, it is particular simple since we write the combined field redefinition in terms of 32-dimensional spinors.
We find thus an elegant way to resolve these problems such that we can compute the final complete Hamiltonian. The cubic and quartic fermionic Hamiltonian, the main results of this paper, are given in Eqs.(\ref{finalH3BF},\ref{finalH4BF}).

With respect to (d), our results are that we find expressions for the Green-Schwarz
superstring Lagrangian and Hamiltonian in the full 32 dimensional
spinor space and that we derive a consistent light-cone gauge fixing for
states belonging to $\C P^3$.
The quadratic part of the fermionic
action of~\cite{Cvetic:1999zs}  is sufficient to make one-loop
computations around configurations as folded spinning string in
$\ads_4\times\C P^3$~\cite{McLoughlin:2008he}, but for a complete
quantum calculation of the finite-size corrections to the spectrum of strings
states, the full Hamiltonian is needed including all terms that are quadratic, cubic
and quartic in the number of fields.

With respect to (e), our paper obviously builds on early papers
\cite{Astolfi:2008ji,Sundin:2008vt} where the bosonic string
spectrum was examined in the $SU(2)\times SU(2)$ sector using only
the bosonic Hamiltonian by employing zeta-function regularization to
regularize divergences coming from the cubic Hamiltonian at
the second order in the perturbative expansion in the inverse of the curvature radius $R$. In
those papers it was assumed that the four-field terms in the
Hamiltonian are normal ordered. However, this seems on further
scrutiny not a valid assumption since the divergent contributions
coming from second order perturbation theory  are, on general
grounds, always negative and cannot be canceled by analogous terms
coming from the cubic Hamiltonian with two fermion and one boson
fields~\cite{futurework}. Thus, one should include all the non-normal-ordered terms
also in the four bosons, four fermions and two-fermion-two-boson parts of the full Hamiltonian to obtain the correct
spectrum~\footnote{These crucial terms were ignored in the analysis
of ref.~\cite{Sundin:2009zu}}. 



The explicit  construction of  the complete $\ads_4\times \C P^3$
sigma model including all the 32 Grassmann--odd coordinates was done
in refs.~\cite{Gomis:2008jt,Grassi:2009yj}. Whereas in the
maximally supersymmetric $\ads_5\times S^5$ background the
supergeometry is described by the coset superspace $SU(2, 2|4)/SO(5)
\times SO(1, 4)$ one instead has that  the type IIA $\ads_4\times \C P^3$ superspace is
not a coset superspace. Its supergeometry can
be completely characterized by the $OSp(6|4)/U(3)\times SO(1,3)$
coset superspace only on a submanifold of the superspace~\cite{Arutyunov:2008if,Stefanski:2008ik}. On this
submanifold the classical superstring equations of motion are
integrable~\cite{Arutyunov:2008if,Stefanski:2008ik},  generalizing
the corresponding result for
type IIB superstring propagating on the $\ads_5\times S^5$
supercoset~\cite{Bena:2003wd}. We are using the type IIA Green-Schwarz action of
refs.~\cite{Gomis:2008jt,Grassi:2009yj} on this particular submanifold.
However, in the $\ads_4\times \C P^3$
superspace there is a different submanifold described by a ``twisted"
$OSp(2|4)/SO(2)\times SO(1,3)$ superspace, which is not a
supercoset, and the ingredients used to prove integrability in
\cite{Bena:2003wd} do not directly apply to this sector of the
theory. Therefore, it remains an open problem to determine whether
the complete set of classical equations of motion of the
Green-Schwarz superstring propagating on the $\ads_4\times \C P^3$
superspace is even classically integrable. The fact that the
$\ads_4\times \C P^3$ superspace with 32 fermionic directions is not
a supercoset requires in fact more general techniques to prove
classical integrability.

Several papers have developed the Lagrangian (and in some cases the
Hamiltonian) for the superspace construction by using a 24
dimensional spinor space that is manifestly supersymmetric
\cite{Arutyunov:2008if,Stefanski:2008ik,Fre:2008qc,Rashkov:2008rm,Dukalski:2009pr,Sundin:2009zu}. In particular in
\cite{Sundin:2009zu} the four-fermion Hamiltonian is found, however,
the complete quartic Hamiltonian has not been computed, only
preliminary versions of the interacting Lagrangian and Hamiltonian have been
provided. In order to have a complete
characterization of the spectrum, it is necessary to derive these
objects carefully dealing with all the issues we described above. This is what we do in this paper.

Our main motivations for this work come from some interesting
differences between the
 $\ads_4 / \mbox{CFT}_3$ duality and the $\ads_5 / \mbox{CFT}_4$ duality:
\begin{itemize}
\item[(1)] The magnon dispersion relation in the $\ads_4 / \mbox{CFT}_3$ duality
can vary as a function of $\lambda$. Indeed, shortly after the
discovery of the $\ads_4 / \mbox{CFT}_3$ duality it was found that a
magnon in the $SU(2)\times SU(2)$ sector of ABJM theory has a
dispersion relation that depends non-trivially on the coupling
\cite{Gaiotto:2008cg,Grignani:2008is,Nishioka:2008gz}
\begin{equation}
\label{dispgen} \Delta = \sqrt{ \frac{1}{4} + h(\lambda) \sin^2
\Big( \frac{p}{2} \Big) } \spa h(\lambda) = \left\{ \begin{array}{c}
\ds 4\lambda^2 + \CO ( \lambda^4 )
 \ \mbox{for} \ \lambda \ll 1 \\[4mm] \ds 2 \lambda + \CO ( \sqrt{\lambda} )
 \ \mbox{for} \ \lambda \gg 1 \end{array} \right.
\end{equation}
where the weak coupling result is from
\cite{Minahan:2008hf,Gaiotto:2008cg}. Corrections to the leading
weak and strong coupling results have been discussed in~\cite{Kristjansen:2008ib,Zwiebel:2009vb,Bak:2009mq,McLoughlin:2008ms,Alday:2008ut,Krishnan:2008zs,McLoughlin:2008he,Bandres:2009kw}.
%
%
\item[(2)] In the $\ads_4 / \mbox{CFT}_3$ duality one has $4_B + 4_F$ magnons, $i.e.$
four bosonic and four fermionic magnons, in the Bethe ansatz for
ABJM theory. However, the pp-wave background has $8_B+8_F$ magnons.
$4_B + 4_F$ of these (the light magnons) correspond to the $4_B+4_F$
magnons in the Bethe ansatz. The other $4_B+4_F$ magnons (the heavy
magnons) should instead somehow emerge from the spectrum of the
light magnons. This is discussed in \cite{Zarembo:2009au,
Sundin:2009zu}.
\item[(3)] While the $\CN =4$ SYM theory and the $\ads_5\times S^5$
background have the maximally possible amount of supersymmetries
with 32 supercharges preserved, the $\CN=6$ superconformal
Chern-Simons theory of ABJM and the $\ads_4 \times \C P^3$
background have 24 supercharges preserved. This allows for the
radius of $\ads_4 \times \C P^3$ to vary as a function of $\lambda$
\cite{Bergman:2009zh}.
\end{itemize}

With respect to both $(1)$ and $(2)$ it is very important to find
the leading finite-size corrections to the quantum string spectrum.
Finding the quantum string spectrum including in particular also the
non-normal-ordered terms in the Hamiltonian should yield a finite
spectrum for the quantum string without need of regularizing. With
respect to $(1)$ this will settle the issue of what the
$1/\sqrt{\lambda}$ correction to $h(\lambda)$ is for large
$\lambda$. This is an important question since it has been found
that there are certain semi-classical spinning string configurations
for which the one-loop correction to the leading energy can only
match with the all-loop Bethe ansatz of
\cite{McLoughlin:2008ms,Alday:2008ut,Krishnan:2008zs,McLoughlin:2008he,Bandres:2009kw}
provided there is a certain non-zero value for this
$1/\sqrt{\lambda}$ correction. Instead other calculations
\cite{Gromov:2008bz,Gromov:2008fy} have found that this correction
should be zero. It is even speculated if this number is measurable,
or if it is scheme dependent, since one can make redefinitions of
the coupling. However, if the quantum string spectrum reveals the
same answer for this correction as the one-loop correction to the
semi-classical string configuration it would suggest that string
theory picks out a unique value. For the spinning string this matter has been thoroughly discussed in~\cite{McLoughlin:2008he}.  We postpone the computation of the
quantum string spectrum to a later publication \cite{futurework}.

With respect to $(2)$, it would be important to examine how the
heavy $4_B+4_F$ magnons in the pp-wave background can emerge from
the $4_B+4_F$ light magnons in the Bethe ansatz. A proposal for how
this works in the continuum limit is presented in
\cite{Zarembo:2009au}. However, it is not immediately clear how this
proposal should resolve the problem for the discrete spectrum of the
quantum string. Certainly, the completely well-defined and explicit expressions for the
full interacting Hamiltonian that we provide in this paper, allowing for a complete calculation of the oscillator spectrum, will shed some light on this subtle problem.

This paper is built up as follows. In Section \ref{sec:genlagr}  we compute the Lagrangian for the type IIA Green-Schwarz (GS) superstring in $\ads_4 \times \C P^3$ using the superspace construction of \cite{Gomis:2008jt,Grassi:2009yj}. In Section~\ref{sec:equi} we prove the equivalence to the general type IIA two-fermion Lagrangian of ref.~\cite{Cvetic:1999zs}. In Section \ref{sec:lclagr}, we analyze the light-cone gauge and the corresponding fixing of $\kappa$-symmetry.  In Section \ref{sec:ppwave} we find the pp-wave Lagrangian and Hamiltonian and derive the pp-wave spectrum for the light and heavy modes. In Section \ref{sec:lagr_redef} we provide the field redefinitions on the Lagrangian necessary to pass to the Hamiltonian formalism. Finally Section~\ref{sec:lcham} contains our results for the full light-cone Hamiltonian up to terms quartic in the number of fields.
Appendix \ref{app:background} contains the details of the $\ads_4 \times \C P^3$ background along with the near plane wave limit. Our conventions for the Gamma-matrices are instead given in Appendix \ref{app:gamma}.  In Appendix \ref{app:Msqr} we derive the structure constants of the $OSp(6|2,2)$ algebra and the fermionic matrix entering in the four fermion terms of the Lagrangian.

\section{Lagrangian for type IIA superstring from superspace}
\label{sec:genlagr}

In this section we present the Lagrangian for the type IIA Green-Schwarz (GS) superstring in $\ads_4 \times \C P^3$ using the superspace construction of~\cite{Gomis:2008jt,Grassi:2009yj}. We restrict ourselves to the supersymmetric fermionic directions. We consider the light-cone gauge and the corresponding fixing of $\kappa$-symmetry in Section \ref{sec:lclagr}. The $\ads_4 \times \C P^3$ background is presented in Appendix \ref{app:background}. The Gamma-matrix conventions are presented in Appendix \ref{app:gamma}.

\subsection{Supersymmetric fermionic directions}

For the type IIA GS string we have two Majorana-Weyl spinors $\theta^{1,2}$ with opposite chirality, $i.e.$ $\Gamma_{11} \theta^1 = \theta^1$ and $\Gamma_{11} \theta^2 = - \theta^2$. We collect these into a 32 component real spinor $\theta = \theta^1 + \theta^2$.

In the superspace construction of \cite{Gomis:2008jt,Grassi:2009yj} all the 32 real fermionic directions of $\theta$ are considered. 24 of these are supersymmetric and 8 are non-supersymmetric. We shall restrict ourselves to the 24 directions which are supersymmetric since we are interested in considering curvature corrections the pp-wave background that comes from a Penrose limit corresponding to a null geodesic moving on an isometry of $\C P^3$. This means that we can choose a gauge for the $\kappa$-symmetry of the type IIA GS Lagrangian where the 8 non-supersymmetric fermionic directions are put to zero \cite{Arutyunov:2008if,Gomis:2008jt,Grassi:2009yj}.

The 24 supersymmetric directions are given as follows. Recall the matrix $J$ of Eq.~(\ref{defQ})
\begin{equation}
\label{defJ} J = \Gamma_{0123} \Gamma_{11} ( - \Gamma_{49} -
\Gamma_{56} + \Gamma_{78} ) = \Gamma_{5678} - \Gamma_{49}
(\Gamma_{56} - \Gamma_{78})
\end{equation}
Note that $J^T = J$ and $J^2 = 2J + 3$.
The matrix $J$ is defined such that it is proportional to $F_{ab} \Gamma^{ab}$ where $F_{(2)}$ is the two-form field strength given in \eqref{F2bein}, $a,b=0,...,9$ being flat target space-time indices. $F_{(2)}$ is proportional to the K\"ahler form on $\C P^3$. In terms of \eqref{defJ}
 the projector on to the supersymmetric fermionic directions is
\begin{equation}
\label{defP} P = \frac{3-J}{4}
\end{equation}
Thus all supersymmetric fermionic directions are characterized by $P \theta = \theta$ or equivalently $J \theta = - \theta$.

If $\Gamma_{5678} \theta = - \theta$ we see that $J \theta = - \theta$. Hence this gives 16 supersymmetric directions. If $\Gamma_{5678} \theta = \theta$ then we need in addition that $\Gamma_{4956} \theta =\theta$. Hence this gives 8 supersymmetric directions. Thus we see that we have in total 24 supersymmetric directions. The 8 remaining non-supersymmetric directions are characterized by $\Gamma_{5678} \theta = \theta$ and $\Gamma_{4956} \theta =-\theta$ corresponding to $J \theta = 3\theta$.

\subsection{Supervielbeins and Lagrangian}

We now present the GS Lagrangian for the 24 supersymmetric directions.
Thus, we assume in the following that $\theta$ obey $P \theta=\theta$.

Write the world-sheet metric as $s_{AB}$ with the world-sheet indices $A,B=0,1$. Then we
define $h^{AB} = \sqrt{|\det s |} s^{AB}$. Thus $\det h = -1$. We
furthermore define the epsilon symbol $\varepsilon^{AB}$ such that
$\varepsilon^{01} = \varepsilon_{01} = 1$.

Introduce for $0 \leq s \leq 1$ the supervielbeins
\begin{equation}
\label{superviel1}
E(s)^a = e^a +  4 i \bar{\theta} \Gamma^a \frac{\sinh^2 (
\frac{s}{2} \CM )}{\CM^2} D \theta \spa E(s)^\alpha = \left(
\frac{\sinh s \CM }{\CM} D \theta \right)^\alpha
\end{equation}
where $a$ is the flat target space-time index. We write
\begin{equation}
\label{superviel2}
E^a = E(s=1)^a = e^a +  4 i \bar{\theta} \Gamma^a \frac{\sinh^2 (
\frac{1}{2} \CM )}{\CM^2} D \theta \spa E^\alpha = E(s=1)^\alpha =
\left( \frac{\sinh \CM }{\CM} D \theta \right)^\alpha
\end{equation}
The covariant derivative is
\begin{equation}
\label{covdertheta}
D \theta = P ( d - \frac{1}{R} \Gamma_{0123} \Gamma_a e^a +
\frac{1}{4} \omega^{ab} \Gamma_{ab} ) \theta
\end{equation}
The two-fermion matrix $\CM^2$ can be found in terms of the structure constants of the generators of $OSp(6|2,2)$. Schematically we write
\begin{equation}
(\CM^2)^\alpha_\beta = - \theta^\gamma \tilde{f}^\alpha_{\gamma i} \theta^\delta \hat{f}^i_{\delta \beta}
\end{equation}
in terms of the structure constants of the $OSp(6|2,2)$ algebra \eqref{fullalg} given explicitly by \eqref{theftildes}-\eqref{thefhats} in Appendix \ref{app:Msqr}. By Eq.~\eqref{them2formula} in Appendix \ref{app:Msqr} we have
\begin{eqnarray}
\left(\mathcal{M}^2\right)_\beta^\alpha &=&-\frac{2i}{R}(P\Gamma_{0123}\Gamma_{\hat{a}}
P)^{\alpha}_{\ \gamma}\theta^{\gamma}\theta^{\delta}(P\Gamma^{0}\Gamma^{\hat{a}} P)_{\delta
\beta} - \frac{i}{R}(P\Gamma_{0123}\Gamma_{{a'}} P)^{\alpha}_{\
\gamma}\theta^{\gamma}\theta^{\delta}(P\Gamma^{0}\Gamma^{{a'}} P)_{\delta
\beta}\cr&+&\frac{i}{R}(P\Gamma_{11}\Gamma_{{a'}} P)^{\alpha}_{\
\gamma}\theta^{\gamma}\theta^{\delta}(P\Gamma^{0}\Gamma_{0123}\Gamma_{11}\Gamma^{{a'}} P)_{\delta
\beta} - \frac{i}{R}(P\Gamma_{\hat{a}\hat{b}} P)^{\alpha}_{\
\gamma}\theta^{\gamma}\theta^{\delta}(P\Gamma^{0}\Gamma_{0123}\Gamma^{\hat{a}\hat{b}} P)_{\delta
\beta}\cr &+&\frac{i}{2R}(P\Gamma_{{a'}{b'}} P)^{\alpha}_{\
\gamma}\theta^{\gamma}\theta^{\delta}(P\Gamma^{0}\Gamma_{0123}\Gamma^{{a'}{b'}} P)_{\delta
\beta}- \frac{i}{R}(\Gamma_{0123}\Gamma_{11} )^{\alpha}_{\
\gamma}\theta^{\gamma}\theta^{\delta}(\Gamma^{0}\Gamma_{11})_{\delta \beta}\cr&&
\end{eqnarray}
where $\hat{a},\hat{b}=0,1,2,3$ and $a',b'=4,...,9$.
Note that $\CM^2 = P \CM^2 P$.
It is shown in Appendix \ref{app:Msqr} that this expression is equivalent to the one found in \cite{Gomis:2008jt,Grassi:2009yj} in a different representation of the $OSp(6|2,2)$ algebra.

From the supervielbeins \eqref{superviel1}-\eqref{superviel2} we can construct the generalized Maurer-Cartan forms
\begin{equation}
L(s)_{A}^a = E(s)^a_\mu \partial_A X^\mu  + E(s)_\alpha^a \partial_A
\theta^\alpha  \spa L(s)^\alpha_A = E(s)^\alpha_\mu \partial_A X^\mu
+ E(s)_\beta^\alpha \partial_A \theta^\beta
\end{equation}
for $0 \leq s \leq 1$. We define then the Maurer-Cartan forms $L^a_A = L(s=1)^a_A$ and $L^\alpha_A = L(s=1)^\alpha_A$. We can now write the type IIA GS Lagrangian for the supersymmetric fermionic directions on $\ads_4\times \C P^3$ \cite{Gomis:2008jt,Grassi:2009yj}, based on the supercoset construction of \cite{Metsaev:1998it,Kallosh:1998zx}, as
\begin{equation}
\label{generallagr}
\CL = - \frac{1}{2} h^{AB} \eta_{ab} L^a_A L^b_B - 2i \varepsilon^{AB}
\int_0^1 ds L(s)^a_{A} ( \bar{\theta} \Gamma_a \Gamma_{11} )_\alpha
L(s)^\alpha_{B}
\end{equation}
The Virasoro constraints are
\begin{equation}
\label{generalvir}
S_{AB} = \frac{1}{2} h_{AB} h^{CD} S_{CD} \spa S_{AB} \equiv
\eta_{ab} L^a_A L^b_B
\end{equation}

From now on we shall truncate the Lagrangian to include terms with at most four fermions, since this is the order relevant to compute one-loop corrections to pp-wave energies.
Dividing $S_{AB}$ according to the number of fermions we have
\begin{equation}
S_{AB} = S_{AB}^{\rm (0 f)} +S_{AB}^{\rm (2 f)} +S_{AB}^{\rm (4 f)}
\end{equation}
we compute
\begin{equation}
S_{AB}^{\rm (0 f)} = g_{\mu\nu} \partial_A X^\mu
\partial_B X^\nu
\end{equation}
\begin{equation}
\label{SAB2f}
S_{AB}^{\rm (2 f)} = i \bar{\theta} \Gamma_\mu (
\partial_A X^\mu D_B \theta + \partial_B X^\mu D_A
\theta )
\end{equation}
\begin{equation}
\label{SAB4F} S_{AB}^{\rm (4 f)} = - ( \bar{\theta} \Gamma^a D_A
\theta )( \bar{\theta} \Gamma_a D_B \theta ) + \frac{i}{12}
\bar{\theta} \Gamma_\mu \CM^2 (\partial_A X^\mu D_B \theta +
\partial_B X^\mu D_A \theta)
\end{equation}

We write the total Lagrangian \eqref{generallagr} as
\begin{equation}
\CL = \CL_{\rm kin} + \CL_{\rm WZ}
\end{equation}
where the kinetic part is
\begin{equation}
\CL_{\rm kin} = - \frac{1}{2} h^{AB} S_{AB}
\end{equation}
and the Wess-Zumino part is
\begin{equation}
\CL_{\rm WZ} = \CL_{\rm WZ}^{\rm (2 f)} + \CL_{\rm WZ}^{\rm (4 f)}
\end{equation}
where
\begin{equation}
\label{LWZ2F} \CL_{\rm WZ}^{\rm (2 f)} =  - i \varepsilon^{A B}
\partial_A X^\mu  \bar{\theta} \Gamma_\mu \Gamma_{11} D_B
\theta
\end{equation}
and
\begin{equation}
\label{LWZ4F} \CL_{\rm WZ}^{\rm (4 f)} = - \frac{i}{12} \varepsilon^{A
B}
\partial_A X^\mu \bar{\theta} \Gamma_\mu \Gamma_{11}
\CM^2 D_B \theta  + \frac{1}{2} \varepsilon^{A B} (\bar{\theta}
\Gamma^a D_A \theta)  (\bar{\theta} \Gamma_a \Gamma_{11} D_B \theta)
\end{equation}

\section{Equivalence with general type IIA Lagrangian for two-fermion terms}
\label{sec:equi}

In this section we show that the two-fermion terms in the type IIA superspace Lagrangian \eqref{generallagr} for the $\ads_4 \times \C P^3$ background, restricting to the supersymmetric fermionic directions, are equivalent to those of the type IIA GS Lagrangian, found by Cvetic et al.~\cite{Cvetic:1999zs} for general type IIA backgrounds, on this particular background. We use here the explicit expressions for the $\ads_4 \times \C P^3$ background written in Appendix \ref{app:background}.%
\footnote{While this paper was in preparation the paper \cite{Cagnazzo:2009zh} appeared where the equivalence between \eqref{generallagr} and the Lagrangian of \cite{Cvetic:1999zs} is also examined.}

\subsubsection*{General GS superstring action for type IIA}

The type IIA superstring Lagrangian including two-fermion terms for
a background with zero Kalb-Ramond field and zero dilaton field can be written as for the superspace, as  a sum of kinetic and Wess-Zumino part
\cite{Cvetic:1999zs}
\begin{equation}
\label{cvet1}
\CL = \CL_{\rm kin } + \CL_{\rm WZ}
\end{equation}
with the kinetic part given by
\begin{equation}
\label{cvet2}
\CL_{\rm kin} = - \frac{1}{2} h^{AB} S_{AB}
\end{equation}
\begin{equation}
\label{cvetSAB} S_{AB}  =  g_{\mu\nu} \partial_A X^\mu
\partial_B X^\nu + i \bar{\theta} \Gamma_\mu (
\partial_A X^\mu \tilde{D}_B \theta +
\partial_B X^\mu \tilde{D}_A \theta )  + \frac{i}{8}  \partial_A X^\mu
\partial_B X^\nu  \bar{\theta} ( \Gamma_\mu M \Gamma_\nu  +  \Gamma_\nu M
\Gamma_\mu ) \theta
\end{equation}
and the Wess-Zumino part given by
\begin{equation}
\label{cvetWZ} \CL_{\rm WZ}  =    i \varepsilon^{AB}  \bar{\theta}
\Gamma_{11}\Gamma_\mu
\partial_A X^\mu \tilde{D}_B \theta  +  \frac{i}{8} \varepsilon^{AB}  \partial_A X^\mu
\partial_B X^\nu   \bar{\theta} \Gamma_{11} \Gamma_\mu M \Gamma_\nu \theta
\end{equation}
where the matrix $M$ is defined as
\begin{equation}\label{M}
M = -  \frac{1}{2} F_{\mu\nu} \Gamma_{11} \Gamma^{\mu\nu} +
\frac{1}{24} F_{\mu\nu\rho\sigma} \Gamma^{\mu\nu\rho\sigma}
\end{equation}
where $F_{(2)}$ and $F_{(4)}$ are the two and four-form
Ramond-Ramond field strengths and the covariant derivative $\tilde{D}_A \theta$  is
\begin{equation}\label{covder}
\tilde{D}_A \theta = \partial_A \theta + \frac{1}{4} \partial_A X^\mu \omega^{ab}_\mu
\Gamma_{ab} \theta
\end{equation}
where  ${\omega_\mu}^{ab}$ is the spin-connection with $a,b$ being
flat indices.
The Virasoro constraints are again given by
\begin{equation}
\label{cvetvir}
S_{AB} = \frac{1}{2} h_{AB} h^{CD} S_{CD}
\end{equation}

\subsubsection*{Equivalence with superspace action}

Using \eqref{F2bein}-\eqref{F4bein} we compute
\begin{equation}
M = - \frac{8}{R} \Gamma_{0123} P
\end{equation}
Inserting this into \eqref{cvetSAB} and \eqref{cvetWZ}, with covariant derivative \eqref{covder}, we see that Eqs.~\eqref{cvetSAB} and \eqref{cvetWZ} are equivalent to Eqs.~\eqref{SAB2f} and \eqref{LWZ2F}, with covariant derivative \eqref{cvetvir}, provided we have that $\tilde{D}_A \theta = P \tilde{D}_A \theta$
for any spinor with $P \theta = \theta$ on the $\ads_4\times \C P^3$
background. This is true if
\begin{equation}
\label{Prelation} \omega^{ab}_\mu [ P , \Gamma_{ab} ] = 0
\end{equation}
We can now check this using the spin connection as computed from the
zehnbeins \eqref{zehnbeins1}-\eqref{zehnbeins4}. For nearly all
non-zero components of $\omega_\mu^{ab}$ you have that $a$ and $b$
are such that $[P,\Gamma_{ab}]=0$. The only 8 components for which
this is not the case are $\omega_{x_1}^{45}$, $\omega_{x_1}^{69}$,
$\omega_{y_1}^{46}$, $\omega_{y_1}^{59}$, $\omega_{x_2}^{47}$,
$\omega_{x_2}^{89}$, $\omega_{y_2}^{48}$ and $\omega_{y_2}^{79}$.
However, if we consider $\omega_{x_1}^{45}$ and $\omega_{x_1}^{69}$
we see that $\omega_{x_1}^{45} = - \omega_{x_1}^{69}$. For
\eqref{Prelation} to hold for $\mu=x_1$ it is therefore sufficient
that $[P, \Gamma_{45}-\Gamma_{69}]=0$ which indeed is the case, as
one can check explicitly using \eqref{defJ}-\eqref{defP}. It works
similar for $\mu=y_1,x_2,y_2$ hence we have checked explicitly that
Eq.~\eqref{Prelation} holds. Actually the group theoretical reason of ~\eqref{Prelation} is as follows. The $\C P^3$ part of the spin connection
$\omega^{ab} \Gamma_{ab}$ takes values in the algebra of the $SU(3) \times U(1)$ stability group of $\C P^3$. The $U(1)$ subgroup of this stability group is generated by the K\"ahler form $J$ which enters the projector $P$. This insures that $P$ commutes with $\omega^{ab} \Gamma_{ab}$.

We can thus conclude that the two-fermion terms in the Lagrangian and Virasoro constraints \eqref{generallagr}-\eqref{generalvir} agree with the Lagrangian and Virasoro constraints \eqref{cvet1}-\eqref{cvetvir} on the $\ads_4\times \C P^3$ background for the supersymmetric fermionic directions.

\section{Light-cone Lagrangian}
\label{sec:lclagr}

In this section we impose the light-cone gauge for the type IIA GS Lagrangian \eqref{generallagr} on $\ads_4\times \C P^3$ restricted to the supersymmetric fermionic directions. We find explicit expressions for the full gauge-fixed light-cone Lagrangian for terms quadratic, cubic and quartic in the number of fields, expanding in powers of $1/R$ around the pp-wave background where $R$ is the radius of $\C P^3$ in the $\ads_4\times \C P^3$ background (see Appendix \ref{app:background}). The gauge fixed Lagrangian found in this Section is then simplified using the fermionic field redefinition in Section \ref{sec:lagr_redef}.

\subsection{Outline of general procedure}

We consider the $\ads_4 \times  \C P^3 $ background described in Appendix \ref{app:background}. We examine string excitations around a null curve $\delta = t/2$ at $u_i = 0$, $i=1,2,3,4$ and $y_1=y_2=0$ in the limit of $R \rightarrow \infty$. More specifically, we take the near plane wave limit $R\rightarrow \infty$, keeping the coordinates $t,v,u_i,x_a,y_a$ fixed, $i=1,2,3,4$ and $a=1,2$.

In the following we wish to consider the Lagrangian in the light-cone gauge
\begin{equation}
\label{lc1}
t(\tau,\sigma) = c\tau
\end{equation}
\begin{equation}
\label{lc2}
\frac{\partial \CL}{\partial \dot{v}} = \mbox{constant}
\spa \frac{\partial \CL}{\partial v' } = 0
\end{equation}
For choices of the light-cone directions lying both in the $\ads$, see \cite{Uvarov:2009nk}.
We allow for corrections of the world-sheet metric
\begin{equation}
\label{corrwsm}
h^{\tau\tau} = -1 + \frac{q_1}{R} + \frac{q_2}{R^2} + \CO(R^{-3})
\spa h^{\tau\sigma} = \frac{q_3}{R} + \frac{q_4}{R^2} + \CO(R^{-3})
\end{equation}

The procedure is now to construct $S_{AB}$ and $\CL_{\rm WZ}$ from
the expressions of Section \ref{sec:genlagr}. Using $S_{AB}$ with \eqref{lc1}
imposed we can write down the two independent Virasoro constraints.
These two constraints can be solved for $\dot{v}$ and $v'$, order by
order in $1/R$.

Inserting $\dot{v}$ and $v'$ into the two gauge conditions
\eqref{lc2} we can solve for the corrections to the world-sheet
metric \eqref{corrwsm}.

Inserting now both $\dot{v}$ and $v'$ and the corrections to the
world-sheet metric in the following expression
\begin{equation}
\CL_{\rm gf} = \CL - \frac{\partial \CL}{\partial \dot{v}} \dot{v}
\end{equation}
we obtain the gauge fixed Lagrangian $\CL_{\rm gf}$. We write
the expanded gauge fixed Lagrangian as
\begin{equation}\label{lgf}
\CL_{\rm gf} = \CL_{2,B} + \CL_{2,F} + \frac{1}{R} ( \CL_{3,B} +
\CL_{3,BF} ) + \frac{1}{R^2} ( \CL_{4,B} + \CL_{4,BF} + \CL_{4,F} )
+ \CO(R^{-3})
\end{equation}

\subsection{Bosonic terms in Lagrangian}

Employing the procedure explained above it is straightforward to
compute the bosonic part of the Lagrangian
\begin{equation}
\label{CL2B}
\CL_{2,B} = \frac{1}{2} \sum_{i=1}^4 ( \dot{u}_i^2 - {u_i'}^2 - c^2
u_i^2 ) + \frac{1}{16} \sum_{a=1}^2 ( \dot{x}_a^2 - {x_a'}^2 + 2 c
y_a \dot{x}_a + \dot{y}_a^2 - {y_a'}^2 )
\end{equation}
\begin{equation}
\label{CL3B}
\CL_{3,B} = \frac{u_4}{8} ( {x_1'}^2 + {y_1'}^2 - {x_2'}^2 -
{y_2'}^2 - \dot{x}_1^2 - \dot{y}_1^2 + \dot{x}_2^2 + \dot{y}_2^2 )
\end{equation}
\begin{eqnarray}
\label{CL4B} &&\CL_{4,B} = -\frac{2}{c^2}(\sum_{i=1}^8 \dot X^i
{X'}^i)^2-c^2 (\sum_{i=1}^3
u_i^2)^2+\frac{2}{3}c^2u_4^4+\sum_{i,j=1}^3 u_i^2(\dot
u_j^2-{u_j'}^2)\cr && + \frac{1}{2 c^2}(c^2 \sum_{i=1}^4
u_i^2-\sum_{i=1}^8[(\dot X^i)^2+({X'}^i)^2] ) (c^2 \sum_{j=1}^3
u_j^2-3 c^2 u_4^2-\sum_{j=1}^8[(\dot X^j)^2+({X'}^j)^2])\cr &&
-\frac{c}{48}\sum_{a=1}^2 \dot x_a y_a^3 +\frac{1}{16}\sum_{a=1}^2
y_a^2 ({x_a'}^2-\dot x_a^2)
\end{eqnarray}
where we have introduced the compact notation
\begin{equation}\label{coordinates}
\begin{array}{c}
\ds \dot{X}^{i=1...4} = (\dot{u}_1,\dot{u}_2,\dot{u}_3,\dot{u}_4)
\spa \dot{X}^{i=5...8} = \frac{\sqrt{2}}{4}
(\dot{x}_1,\dot{y}_1,\dot{x}_2,\dot{y}_2) \\ \ds {X'}^{i=1...4} =
(u_1',u_2',u_3',u_4') \spa {X'}^{i=5...8} = \frac{\sqrt{2}}{4} (
x_1',y_1',x_2',y_2' )
\end{array}
\end{equation}
%

\subsection{Introduction of fermionic terms}

We define the purely fermionic bilinears
\begin{equation}
\label{ABCdef}
\begin{array}{c} \ds
A_{a,A} = \bar{\theta} \Gamma_a \partial_A \theta \spa
\tilde{A}_{a,A} = \bar{\theta} \Gamma_{11} \Gamma_a \partial_A
\theta \\[2mm] \ds B_{abc} = \bar{\theta} \Gamma_a \Gamma_{bc} \theta
\spa \tilde{B}_{abc} = \bar{\theta} \Gamma_{11} \Gamma_a \Gamma_{bc}
\theta
\\[2mm] \ds
C_{ab} = \bar{\theta}  \Gamma_a P \Gamma_{0123}
\Gamma_b  \theta \spa
\tilde{C}_{ab} =  \bar{\theta} \Gamma_{11}  \Gamma_a P
\Gamma_{0123} \Gamma_b  \theta
\end{array}
\end{equation}
Note the following important properties
\begin{equation}
\begin{array}{c} \ds
C_{ab} = C_{ba} \spa \tilde{C}_{ab} = - \tilde{C}_{ba}
\\[2mm] \ds
B_{abc} = \bar{\theta} \Gamma_{abc} \theta \spa \tilde{B}_{abc} =
\bar{\theta} \Gamma_{11} \Gamma_{abc} \theta
\end{array}
\end{equation}
which can be deduced using that $(\Gamma^0 \Gamma_a)^T = \Gamma^0
\Gamma_a$. In terms of the fermionic bilinears \eqref{ABCdef} the quantities appearing in the
Lagrangian and Virasoro constraints are
\begin{eqnarray}
S^{\rm (2f)}_{AB} &=& i A_{a,B} e^a_\mu \partial_A X^\mu + i A_{a,A}
e^a_\mu \partial_B X^\mu + \frac{i}{4} B_{abc} e^a_\mu
\omega^{bc}_\nu ( \partial_A X^\mu \partial_B X^\nu + \partial_B
X^\mu \partial_A X^\nu ) \nn \\ && - \frac{2i}{R} C_{ab} e^a_\mu
e^b_\nu
\partial_A X^\mu \partial_B X^\nu
\end{eqnarray}
\begin{equation}
\CL_{\rm WZ}^{\rm (2 f)} = i \varepsilon^{AB} \tilde{A}_{a,B} e^a_\mu
\partial_A X^\mu + \frac{i}{4} \varepsilon^{AB} \tilde{B}_{abc}
e^a_\mu \omega^{bc}_\nu \partial_A X^\mu \partial_B X^\nu -
\frac{i}{R} \varepsilon^{AB} \tilde{C}_{ab} e^a_\mu e^b_\nu \partial_A
X^\mu \partial_B X^\nu
\end{equation}
Comparing with Eqs.~\eqref{SAB2f} and \eqref{LWZ2F} we note that
we removed the projector $P$ in our definition of $B_{abc}$ and
$\tilde{B}_{abc}$ in \eqref{ABCdef}. This is allowed since
\eqref{Prelation} holds for the $\ads_4 \times \C P^3$ background.

From \eqref{SAB4F} and \eqref{LWZ4F} we see that the four-fermion terms
are
\begin{eqnarray}
S^{\rm (4f)}_{00} &=& - \eta^{ab} \Big( A_{a,\tau} - \frac{c}{2}
C_{+a} + \frac{c}{4} ( B_{a56} + B_{a78} ) \Big) \Big( A_{b,\tau} -
\frac{c}{2} C_{+b} + \frac{c}{4} ( B_{b56} + B_{b78} ) \Big) \nn \\
&& + \frac{ic}{6} \bar{\theta} \Gamma_+ \CM^2 \dot{\theta} -
\frac{ic^2}{12} \bar{\theta} \Gamma_+ \CM^2 P \Gamma_{0123} \Gamma_+
\theta + \frac{ic^2}{24} \bar{\theta} \Gamma_+ \CM^2 (
\Gamma_{56}+\Gamma_{78} ) \theta
\end{eqnarray}
\begin{equation}
S^{\rm (4f)}_{01} = - \eta^{ab} \Big( A_{a,\tau} - \frac{c}{2}
C_{+a} + \frac{c}{4} ( B_{a56} + B_{a78} ) \Big) A_{b,\sigma} +
\frac{ic}{12} \bar{\theta} \Gamma_+ \CM^2 \theta'
\end{equation}
\begin{equation}
S^{\rm (4f)}_{11} = - \eta^{ab} A_{a,\sigma} A_{b,\sigma}
\end{equation}
\begin{eqnarray}
\CL_{\rm WZ}^{\rm (4 f)} &=&  \frac{ic}{12} \bar{\theta} \Gamma_{11}
\Gamma_+ \CM^2 \theta'   - \frac{1}{2} \eta^{ab} \Big[ A_{a,\tau} -
\frac{c}{2} C_{+a} +
\frac{c}{4} ( B_{a56} + B_{a78} ) \Big] \tilde{A}_{b,\sigma}  \nn \\
&& + \frac{1}{2} \eta^{ab} \Big[ \tilde{A}_{a,\tau} + \frac{c}{2}
\tilde{C}_{+a} + \frac{c}{4} ( \tilde{B}_{a56} + \tilde{B}_{a78} )
\Big] A_{b,\sigma}
\end{eqnarray}
where we used that
\begin{equation}
\begin{array}{c} \ds
\bar{\theta} \Gamma_a D_\tau \theta = A_{a,\tau} - \frac{c}{2}
C_{+a} + \frac{c}{4} ( B_{a56} + B_{a78} ) + \CO( R^{-1} ) \\[3mm]
\ds \bar{\theta} \Gamma_{11} \Gamma_a D_\tau \theta =
\tilde{A}_{a,\tau} + \frac{c}{2} \tilde{C}_{+a} + \frac{c}{4} (
\tilde{B}_{a56} + \tilde{B}_{a78} ) + \CO( R^{-1} )
\end{array}
\end{equation}

\subsection{Fixing $\kappa$-symmetry}

We are considering here the $\kappa$-symmetry transformation on the supersymmetric fermionic directions. We have already imposed a partial $\kappa$-symmetry gauge choice by demanding $P \theta = \theta$ thus reducing the number of fermionic directions from 32 to 24. In the following we fix the remaining 8 directions in the $\kappa$-symmetry by a gauge choice that follows from our light-cone gauge.

In general the $\kappa$-symmetry variations are (assuming we are on the space of supersymmetric directions $P \theta = \theta$)
\begin{equation}
\label{kap1}
E^\alpha_\mu \delta X^\mu +  E^\alpha_\beta \delta \theta^\beta = [(1+\Gamma) \kappa]^\alpha
\end{equation}
\begin{equation}
\label{kap2}
E^a_\mu \delta X^\mu  +  E^a_\alpha \delta \theta^\alpha = 0
\end{equation}
where $(1+\Gamma)/2$ is a spinor projection matrix defined in~\cite{Green:1983wt, Green:1983sg, Grisaru:1985fv}.

We analyze the $\kappa$-symmetry in the Penrose limit. In this limit we have the super vielbeins
\begin{equation}
E^a = e^a + i \bar{\theta} \Gamma^a D \theta
\spa
E^\alpha = ( D \theta )^\alpha
\end{equation}
Eqs.~\eqref{kap1}-\eqref{kap2} give
\begin{equation}
\delta X^\mu (D_\mu \theta)^\alpha + \delta \theta^\alpha = [(1+\Gamma) \kappa]^\alpha
\spa
\delta t + 2i   \bar{\theta} \Gamma^+ D_\mu \theta \delta X^\mu + 2i \bar{\theta} \Gamma^+ \delta \theta = 0
\end{equation}
using $e^+ = \frac{1}{2} dt$ and assuming $P\theta=\theta$ and $P\delta \theta=\delta \theta$. Combining these equations we get
\begin{equation}
\label{deltatp}
\delta t = - 2i \bar{\theta} \Gamma^+ ( 1+\Gamma) \kappa
\end{equation}
For the light-cone gauge to be consistent we need that $\delta t =
0$ under variations of $\kappa$-symmetry. Suppose now we have a
supersymmetric fermionic direction with $P \Gamma^- P \theta = 0$.
Then we see from \eqref{deltatp} that $\delta t =  2i \theta^T
(1+\Gamma) \kappa$. Thus, such fermionic directions are clearly not
consistent with the light-cone gauge. The matrix $P \Gamma^- P$ has
8 supersymmetric fermionic directions with eigenvalue zero,
characterized by $\Gamma_{5678} \theta = - \theta$ and $\Gamma^-
\theta = 0$. We fix the remaining $\kappa$-symmetry gauge
freedom by demanding that these directions are put to zero. This
leaves the following 16 physical fermionic directions in the light
cone gauge
\begin{equation}\label{fermgauge}
\begin{array}{c}
\mbox{8 fermionic directions defined by} \ \Gamma_{5678}\theta = -\theta , \ \Gamma^+ \theta = 0 \\
\mbox{8 fermionic directions defined by} \ \Gamma_{5678}\theta =
\theta , \ \Gamma_{4956} \theta = \theta
\end{array}
\end{equation}
It is useful to parameterize this by introducing the projectors
\begin{equation}
\label{ppm} \begin{array}{c} \ds \CP_+ = \frac{I+\Gamma_{5678}}{2}
\frac{I+\Gamma_{4956}}{2} \spa \CP_- = \frac{I-\Gamma_{5678}}{2}
\frac{I-\Gamma_{09}}{2} \\[3mm] \ds \CP_+' = \frac{I+\Gamma_{5678}}{2}
\frac{I-\Gamma_{4956}}{2} \spa \CP_-' = \frac{I-\Gamma_{5678}}{2}
\frac{I+\Gamma_{09}}{2}
\end{array}
\end{equation}
named after the eigenvalue of $\Gamma_{5678}$. These projectors are mutually orthogonal to each other and they are all idempotent and symmmetric. We have
\begin{equation}
P = \CP_+ + \CP_- + \CP_-' \spa I = \CP_+ + \CP_- + \CP_+' + \CP_-'
\end{equation}
We are thus imposing that our spinor $\theta=\theta^1+\theta^2$
obeys
\begin{equation}
\label{physferm} ( \CP_+ + \CP_- ) \theta = \theta
\end{equation}
or, equivalently, $( \CP_+' + \CP_-' )\theta=0$.
This is our condition for physical fermionic modes.

It is worth noting that the gauge condition \eqref{fermgauge}  is different from the $\kappa$-symmetry gauge fixing condition that one imposes for string theory in  $AdS_5 \times S^5$. As it is well known, in that case it is sufficient to impose the condition $ \Gamma^+ \theta = 0$ for all the fermionic directions, while in this case, due to the less amount of supersymmetry, the condition $ \Gamma^+ \theta = 0$ does not only select supersymmetric states.

\subsection{Explicit fermionic terms in Lagrangian}

We compute here the fermionic quantities appearing in (\ref{lgf}) in terms of the fermionic bilinears defined in (\ref{ABCdef})
\begin{equation}
\label{CL2F}
\CL_{2,F} = \frac{ic}{2} A_{+,\tau} + \frac{ic}{2}
\tilde{A}_{+,\sigma} + \frac{ic^2}{8} ( B_{+56}+B_{+78} ) -
\frac{ic^2}{4} C_{++}
\end{equation}
\begin{equation}
\label{CL3BF}
\begin{array}{l} \ds
\CL_{3,BF} = - ic \sum_{i=1}^8 ( \tilde{C}_{+i} {X'}^i + C_{+i}
\dot{X}^i)   + i \sum_{i=1}^8 \Big[ (A_{i,\tau} +
\tilde{A}_{i,\sigma}) \dot{X}^i - (\tilde{A}_{i,\tau}  +
A_{i,\sigma}) {X'}^i \Big]     \\[2mm] \ds  + \frac{ic}{4} \sum_{i=1}^8 \Big[ ( B_{i56}+B_{i78})
\dot{X}^i
- (\tilde{B}_{i56}+\tilde{B}_{i78}) {X'}^i \Big] + \frac{ic^2}{4} ( B_{+56} - B_{+78} ) u_4 \\[2mm] \ds
+ \frac{ic^2}{4} \sum_{i=1}^4 B_{+-i} u_i
  + \frac{i c
}{4} \sum_{i=5}^8 s_i ( B_{+4i} \dot{X}^i + \tilde{B}_{+4i} {X'}^i ) + \frac{ic}{8} \sum_{i,j=5}^8 \epsilon_{ij} (B_{+-i} \dot{X}^j + \tilde{B}_{+-i} {X'}^j)
\end{array}
\end{equation}
\begin{equation}
\begin{array}{l} \ds
\CL_{4,BF} = - i u_4  \sum_{i=5}^8 s_i \Big[  (A_{i,\tau} + \tilde{A}_{i,\sigma}) \dot{X}^i
- (\tilde{A}_{i,\tau} + A_{i,\sigma} ) {X'}^i  \Big]
+ ic u_4^2 A_{+,\tau} + i c (u_1^2+u_2^2+u_3^2) \tilde{A}_{+,\sigma}
 \\[2mm] \ds
+ \frac{i}{c}
\sum_{i=1}^8 \dot{X}^i {X'}^i  ( \tilde{A}_{+,\tau} -
\tilde{A}_{-,\tau} - A_{+,\sigma} - A_{-,\sigma}  )
+ \frac{i}{2c} \sum_{i=1}^8 ( (\dot{X}^i)^2 + ({X'}^i)^2 )
(A_{+,\tau} + A_{-,\tau} - \tilde{A}_{+,\sigma} + \tilde{A}_{-,\sigma} )
 \\[2mm] \ds
+ i \sum_{i,j=1}^8 \Big[  C_{ij} ({X'}^i {X'}^j - \dot{X}^i \dot{X}^j ) + 2  \tilde{C}_{ij} {X'}^i \dot{X}^j \Big] + i c u_4 \sum_{i=5}^8 s_i ( C_{+i} \dot{X}^i + \tilde{C}_{+i} {X'}^i )
 \\[2mm] \ds
- i \sum_{i=1}^8 \dot{X}^i {X'}^i \tilde{C}_{+-} - \frac{ic^2}{2} \sum_{i=1}^4 u_i^2 C_{++} - \frac{i}{2} \sum_{i=1}^8 ( (\dot{X}^i)^2 + ({X'}^i)^2 ) C_{+-}
+ i c \sum_{i,j=1}^3 u_i' u_j \tilde{B}_{+ij}
 \\[2mm] \ds
+ \frac{ic}{2} u_4 \sum_{i=1}^8 \Big[ (B_{i56} - B_{i78}) \dot{X}^i - ( \tilde{B}_{i56}  - \tilde{B}_{i78}){X'}^i \Big]
- \frac{ic}{4} u_4 \sum_{i=5}^8 s_{i} \Big[ (B_{i56}+B_{i78}) \dot{X}^i - ( \tilde{B}_{i56} + \tilde{B}_{i78}){X'}^i \Big]
 \\[2mm] \ds
- \frac{ic}{2} \sum_{i=1}^8 \sum_{j=1}^4 u_j ( B_{-ij} \dot{X}^i - \tilde{B}_{-ij} {X'}^i )
- \frac{i}{2} \sum_{i=1}^8 \sum_{j=5}^8 s_j \Big[ ( \dot{X}^i \dot{X}^j - {X'}^i {X'}^j ) B_{4 i j} + ( \dot{X}^i {X'}^j - {X'}^i \dot{X}^j ) \tilde{B}_{4 i j} \Big]
 \\[2mm] \ds
+ \frac{ic}{2} \sum_{i=1}^3 \sum_{j=4}^8 u_i \Big[ B_{+ij} \dot{X}^j -  \tilde{B}_{+ij} {X'}^j \Big]
+ \frac{i}{8} \sum_{i=1}^8 ( (\dot{X}^i)^2 + ({X'}^i)^2 ) ( B_{-56} + B_{-78} - B_{+56} - B_{+78} )
 \\[2mm] \ds
+\frac{i}{4} \sum_{i=1}^8 \sum_{j,k=5}^8 \epsilon_{jk} \Big[ (B_{+ij} - B_{-ij}) ( \dot{X}^i \dot{X}^k - {X'}^i {X'}^k )  + (\tilde{B}_{+ij}-\tilde{B}_{-ij}) ( \dot{X}^i {X'}^k - {X'}^i \dot{X}^k )  \Big]
 \\[2mm] \ds
- \frac{ic}{8} u_4 \sum_{i,j=5}^8 s_i \epsilon_{ij} ( B_{+i-} \dot{X}^j + \tilde{B}_{+i-} {X'}^j )
+ \frac{ic}{2} u_4 \sum_{i=1}^3 ( B_{+i4} \dot{u}_i - \tilde{B}_{+i4} u_i' )
 \\[2mm] \ds
- \frac{i}{4} \sum_{i=1}^8 \dot{X}^i {X'}^i (\tilde{B}_{+56}+ \tilde{B}_{+78} + \tilde{B}_{-56}+\tilde{B}_{-78} )
- \frac{ic}{4} ( B_{+56} \dot{x}_1 y_1 +  \tilde{B}_{+56} x_1' y_1 +  B_{+78} \dot{x}_2 y_2 +  \tilde{B}_{+78} x_2' y_2 )
\\[2mm] \ds
+ \frac{ic}{4} u_4 \sum_{i=5}^8 ( -  B_{+4i} \dot{X}^i + 3 \tilde{B}_{+4i} {X'}^i )
+ \frac{i c^2}{4} (B_{+56}+B_{+78}) (u_1^2+u_2^2+u_3^2+2u_4^2)
\end{array}
\end{equation}
\begin{eqnarray}
\label{L4F1}
&& \CL_{4,F} = \frac{ic}{12} \bar{\theta} \Gamma_+ \CM^2 \dot{\theta} -
\frac{ic^2}{24} \bar{\theta} \Gamma_+ \CM^2 P \Gamma_{0123} \Gamma_+
\theta + \frac{ic^2}{48} \bar{\theta} \Gamma_+ \CM^2 (
\Gamma_{56}+\Gamma_{78} ) \theta + \frac{ic}{12} \bar{\theta} \Gamma_{11}
\Gamma_+ \CM^2 \theta' \nn \\ && - \frac{1}{2} \sum_{i=1}^8 \Big[ A_{i,\tau} - \frac{c}{2} C_{+i} + \frac{c}{4} (B_{i56}+B_{i78}) \Big]^2 + \frac{1}{2} \sum_{i=1}^8 A_{i,\sigma}^2 \nn \\ && - \frac{1}{2} \sum_{i=1}^8 \tilde{A}_{i,\sigma} \Big[ A_{i,\tau} - \frac{c}{2} C_{+i} + \frac{c}{4} (B_{i56}+B_{i78}) \Big] + \frac{1}{2} \sum_{i=1}^8 A_{i,\sigma} \Big[ \tilde{A}_{i,\tau} + \frac{c}{2} \tilde{C}_{+i} + \frac{c}{4} (\tilde{B}_{i56}+\tilde{B}_{i78}) \Big] \nn \\ &&
- \frac{1}{2} A_{+,\sigma} \Big[ \tilde{A}_{+,\tau} - \frac{1}{2} \tilde{A}_{-,\tau} - \frac{3c}{4} \tilde{C}_{+-} - \frac{c}{4} ( \tilde{B}_{+56} + \tilde{B}_{+78}) - \frac{c}{8} ( \tilde{B}_{-56} + \tilde{B}_{-78}) \Big] \nn \\ &&
+ \frac{1}{2} \tilde{A}_{+,\sigma} \Big[ A_{+,\tau} + \frac{1}{2} A_{-,\tau} - \frac{c}{2} C_{++} - \frac{c}{4} C_{+-} + \frac{c}{4} ( B_{+56} + B_{+78}) + \frac{c}{8} ( B_{-56} + B_{-78}) \Big] \nn \\ && - \frac{1}{4} A_{-,\sigma} \Big[ \tilde{A}_{+,\tau} +  \frac{c}{4} ( \tilde{B}_{+56} + \tilde{B}_{+78}) \Big] - \frac{1}{4} \tilde{A}_{-,\sigma} \Big[ A_{+,\tau} - \frac{c}{2} C_{++} + \frac{c}{4} ( B_{+56} + B_{+78}) \Big] + \nn \\ && + \frac{c}{4} \Big[ A_{+,\tau} - \frac{c}{2} C_{++} + \frac{c}{4} ( B_{+56} + B_{+78}) \Big] ( C_{+-} - C_{++} + B_{+56} + B_{+78} )
\end{eqnarray}
where we used the notation defined in \eqref{coordinates} and
\begin{equation}
s_5 = s_6 = 1 \spa s_7 = s_8 = -1 \spa \epsilon_{56} = -\epsilon_{65} = \epsilon_{78} = - \epsilon_{87}  = 1
\end{equation}
with all other entries of $\epsilon_{ij}$ being zero.

\section{PP-wave Lagrangian, Hamiltonian and field expansions}
\label{sec:ppwave}

We analyze in this section in detail the pp-wave Lagrangian $\CL_2$ and pp-wave Hamiltonian $\CH_2$ emerging in the $R \rightarrow \infty$ limit. We split up the Lagrangian and Hamiltonian in the bosonic and fermionic parts
\begin{equation}
\CL_2 = \CL_{2,B} + \CL_{2,F}
\spa
\CH_2 = \CH_{2,B} + \CH_{2,F}
\end{equation}
and analyze these separately in the following. The pp-wave spectrum on this
background was derived also in~\cite{Sugiyama:2002tf,Hyun:2002wu,Nishioka:2008gz,Gaiotto:2008cg,Grignani:2008is}. The Penrose limit used here was found in \cite{Grignani:2008is}.%
\footnote{Notice that the resulting pp-wave background has
two flat directions which makes it particularly well suited for studying the $SU(2)\times SU(2)$ sector \cite{Grignani:2008is}. The coordinate system of
this pp-wave is similar to the one found by a Penrose limit in
\cite{Bertolini:2002nr} which is particularly well suited to study the
$SU(2)$ sector of $\ads_5\times S^5$, as explained in
\cite{Harmark:2006ta,Harmark:2008gm}. See also \cite{Astolfi:2008yw}.}

\subsection{Bosonic part}

The quadratic bosonic Lagrangian is
\begin{equation}
\CL_{2,B} = \frac{1}{2} \sum_{i=1}^4 ( \dot{u}_i^2 - {u_i'}^2 - c^2
u_i^2 ) + \frac{1}{16} \sum_{a=1}^2 ( \dot{x}_a^2 - {x_a'}^2 + 2 c
y_a \dot{x}_a + \dot{y}_a^2 - {y_a'}^2 )
\end{equation}
The momentum conjugate fields are defined by
\begin{equation}
\Pi_\mu = \frac{\partial \CL}{\partial \dot{x}^\mu}
\end{equation}
We get $\Pi_{x_a} = (\dot{x}_a + c y_a)/8$, $\Pi_{y_a} =
\dot{y}_a/8$ and $\Pi_{u_i} =\dot{u}_i$. The quadratic bosonic
Hamiltonian is
\begin{equation}
\label{CH2B}
c \CH_{2,B} = \frac{1}{16} \sum_{a=1}^2 \Big[  p_{x_a} ^2 +
p_{y_a}^2  +   {x_a'}^2 + {y_a'}^2  \Big] + \frac{1}{2} \sum_{i=1}^4
\Big[ p_{u_i}^2 + {u_i'}^2 + c^2 u_i^2 \Big]
\end{equation}
where for convenience we defined the fields
\begin{equation}
p_{x_a}  \equiv 8 \Pi_{x_a} - c y_a \spa p_{y_a} \equiv 8 \Pi_{y_a}
\spa p_{u_i}  \equiv \Pi_{u_i}
\end{equation}
Notice that these fields are functions of the momenta and position
variables.

The mode expansion for the bosonic fields can be written as
\begin{equation}
\label{zmode} z_a(\tau,\sigma) = 2 \sqrt{2} \, e^{i\frac{ c
\tau}{2}} \sum_{n \in \Z} \frac{1}{\sqrt{\omega_n}} \Big[ a_n^a
e^{-i ( \omega_n \tau - n \sigma ) } -  (\tilde{a}^a)^\dagger_n e^{i
( \omega_n \tau - n \sigma ) } \Big]
\end{equation}
\begin{equation}
u_i (\tau,\sigma ) = i \frac{1}{\sqrt{2}} \sum_{n\in \Z}
\frac{1}{\sqrt{\Omega_n}} \Big[ \hat{a}^i_n e^{-i ( \Omega_n \tau -
n \sigma ) } - (\hat{a}^i_n)^\dagger e^{i ( \Omega_n \tau - n \sigma
) } \Big]
\end{equation}
where
\begin{equation}
\omega_n=\sqrt{\frac{c^2}{4}+n^2} \spa \Omega_n=\sqrt{c^2+n^2}
\end{equation}
 and
we defined $z_a(\tau,\sigma)=x_a(\tau,\sigma)+iy_a(\tau,\sigma)$.
The canonical commutation relations
$
[x_a(\tau,\sigma),\Pi_{x_b}(\tau,\sigma')] = i\delta_{ab} \delta
(\sigma-\sigma')$, $[y_a(\tau,\sigma),\Pi_{y_b}(\tau,\sigma')] =
i\delta_{ab}\delta (\sigma-\sigma')$ and
$[u_i(\tau,\sigma),\Pi_{u_j}(\tau,\sigma')] = i\delta_{ij} \delta
(\sigma-\sigma')$ follow from
\begin{equation}
\label{comrel} [a_m^a,(a_n^b)^\dagger] = \delta_{mn} \delta_{ab}\spa
[\tilde{a}_m^a,(\tilde{a}_n^b)^\dagger] = \delta_{mn}
\delta_{ab}\spa [\hat{a}^i_m,(\hat{a}^j_n)^\dagger] = \delta_{mn}
\delta_{ij}
\end{equation}
Employing the previous relations we obtain the
bosonic free spectrum as
\begin{equation}
c H_{2} = \sum_{i=1}^4 \sum_{n\in \Z} \Omega_n
\hat{N}^i_n+\sum_{a=1}^2\sum_{n\in \Z} \left(\omega_n -
\frac{c}{2}\right) M_n^a +\sum_{a=1}^2\sum_{n\in \Z} \left(\omega_n+
\frac{c}{2}\right) N_n^a \label{penspectrum}
\end{equation}
with the number operators $\hat{N}^i_n = (\hat{a}^i_n)^\dagger
\hat{a}^i_n$, $M_n^a = (a^a)^\dagger_n a^a_n$ and $N_n^a =
(\tilde{a}^a)^\dagger_n \tilde{a}_n^a$, and with the level-matching
condition
\begin{equation}
\label{levelm} \sum_{n\in \Z}n \left[\sum_{i=1}^4
\hat{N}^i_n+\sum_{a=1}^2 \left(M_n^a + N_n^a\right)\right]
 = 0
\end{equation}
A peculiarity of the bosonic spectrum, as well as of the fermionic one, is that it contains four light and four heavy modes~\cite{Nishioka:2008gz, Grignani:2008is}.  The role of the heavy modes has been investigated in~\cite{Zarembo:2009au, Sundin:2009zu}.

\subsection{Fermionic part}

We now consider the fermionic part of the pp-wave Lagrangian.
From \eqref{CL2F} we have the quadratic fermionic Lagrangian
\begin{equation}
\label{CL2Fb}
\CL_{2,F} = \frac{ic}{2} A_{+,\tau} + \frac{ic}{2}
\tilde{A}_{+,\sigma} + \frac{ic^2}{8} ( B_{+56}+B_{+78} ) -
\frac{ic^2}{4} C_{++}
\end{equation}
Using the identities
\begin{equation}
\label{iden1} \CP_+ \Gamma_{09} = \Gamma_{09} \CP_+' \spa \CP_+'
\Gamma_{09} = \Gamma_{09} \CP_+ \spa \CP_- \Gamma_{09} = \Gamma_{09}
\CP_- = - \CP_-
\end{equation}
we compute
\begin{equation}
\label{iden2}
\begin{array}{c} \ds
( \CP_+ + \CP_-) \Gamma^0 \Gamma_+ ( \CP_+ + \CP_-) = \CP_+ + 2\CP_-
\\[2mm] \ds ( \CP_+ + \CP_-) \Gamma^0 \Gamma_+ P \Gamma_{0123} \Gamma_+ ( \CP_+ + \CP_-)
= ( \CP_+ + 4 \CP_- ) \Gamma_{123} \end{array}
\end{equation}
We can thus write the four terms in \eqref{CL2Fb} as
\begin{equation}
\label{fourterms}
\begin{array}{c} \ds
A_{+,\tau} = \theta (\CP_+ + 2\CP_-) \dot{\theta} \spa \tilde{A}_{+,\sigma} = - \theta (\CP_+ + 2\CP_-) \Gamma_{11} \theta' \\[2mm] \ds B_{+56}+B_{+78} = \theta (\CP_+ + 2\CP_-) ( \Gamma_{56}+\Gamma_{78} )\theta \spa C_{++} = \theta ( \CP_+ + 4 \CP_- ) \Gamma_{123} \theta
\end{array}
\end{equation}
where we used
our $\kappa$-symmetry gauge choice $(\CP_+ +
\CP_-)\theta=\theta$.

Instead of parameterizing the fermionic directions in terms of the 32 component real spinor $\theta = \theta^1 + \theta^2$, $\Gamma_{11}\theta^1 = \theta^1$ and $\Gamma_{11}\theta^2 = - \theta^2$, we parameterize the fermionic directions in terms of the complex spinors
\begin{equation}
\label{defpsi}
\psi = \theta^1 + i \Gamma_{049} \theta^2 \spa \psi^* = \theta^1 - i \Gamma_{049} \theta^2
\end{equation}
We see that $\Gamma_{11} \psi = \psi$ hence $\psi$ has 16 complex components. From our choice of $\kappa$-symmetry gauge we have that physical spinors obey $(\CP_+ + \CP_-) \theta = \theta$ with the projectors $\CP_\pm$ defined by \eqref{ppm}.
Since $\Gamma_{049}$ commutes with $\CP_\pm$ we get that the physical spinors $\psi$ obey the condition
\begin{equation}
(\CP_+ + \CP_- ) \psi = \psi
\end{equation}
In the following we split up the spinor as $\psi = \psi_+ + \psi_-$ with
 $\psi_\pm = \CP_\pm \psi$.

Using the above formulas in \eqref{CL2Fb} we get the Lagrangian
\begin{eqnarray}
\label{finalCL2F}
\CL_{2,F} &=& \frac{ic}{2} \left[ \psi_+^* \dot{\psi_+} +2\psi_-^*\dot{\psi_-} - \frac{1}{2}\left(\psi_+ \psi_+' + \psi_+^* {\psi_+^*}'+2\psi_- \psi_-' +2\psi_-^* {\psi_-^*}'\right)    \right]  \nn \\ && + \frac{c^2}{4} \psi_+ \psi_+^* -c^2\psi_-\psi_-^*+ \frac{ic^2}{2} \psi_-\Gamma_{56}  \psi_-^*
\end{eqnarray}
in terms of the physical spinor fields $\psi_\pm$. Note that we added here the total derivative
\begin{equation}
\frac{ic}{4} \partial_\tau \Big( \psi^*_+ \psi_+ + 2 \psi^*_- \psi_- \Big)
\end{equation}
such that there is no $\dot{\psi}^*$ dependence in the Lagrangian.

From the corresponding e.o.m. we get the following mode expansions
\begin{equation}
\psi_{+,\alpha} =  \frac{\sqrt{2\alpha'}}{\sqrt{c}} \sum_{n\in Z} \Big[ f^+_n d_{n,\alpha}
e^{-i ( \omega_n \tau - n \sigma ) } - f^-_n d^\dagger_{n,\alpha}
e^{i ( \omega_n \tau - n \sigma ) } \Big]
\end{equation}
\begin{equation}
\psi_{-,\alpha} =\frac{\sqrt{\alpha'}}{\sqrt{c}}  \big( e^{- \frac{c}{2} \Gamma_{56}
\tau} \big)_{\alpha \beta} \sum_{n\in Z} \Big[ - g^-_n b_{n,\beta}
e^{-i ( \Omega_n \tau - n \sigma ) } + g^+_n b^\dagger_{n,\beta}
e^{i ( \Omega_n \tau - n \sigma ) } \Big]
\end{equation}
with the constants $f^\pm_n$ and $g^\pm_n$ defined by
\begin{equation}
f^\pm_n = \frac{\sqrt{\omega_n+n} \pm
\sqrt{\omega_n-n}}{2\sqrt{\omega_n}} \spa g^\pm_n =
\frac{\sqrt{\Omega_n+n} \pm \sqrt{\Omega_n-n}}{2\sqrt{\Omega_n}}
\end{equation}
and where the oscillators are subject to the conditions
\begin{equation}
\CP_+ d_n = d_n \spa \Gamma_{11} d_n = d_n \spa \CP_- b_n = b_n \spa \Gamma_{11} b_n = b_n
\end{equation}
and obey the anti-commutation relations
\begin{equation}
\{ d_{m,\alpha} , d^\dagger_{n,\beta} \} = \delta_{mn }
(\frac{1+\Gamma_{11}}{2} \CP_+)_{\alpha\beta} \spa \{ b_{m,\alpha} , b^\dagger_{n,\beta} \} =
\delta_{mn } (\frac{1+\Gamma_{11}}{2} \CP_-)_{\alpha\beta}
\end{equation}

From this we obtain
\begin{equation}
\{ \psi_{+,\alpha} (\tau,\sigma) , \psi_{+,\beta}^* (\tau,\sigma') \} = \frac{4\pi \alpha'}{c} (\frac{1+\Gamma_{11}}{2}\CP_+)_{\alpha\beta} \delta ( \sigma-\sigma' )
\end{equation}
\begin{equation}
\{ \psi_{-,\alpha} (\tau,\sigma) , \psi_{-,\beta}^* (\tau,\sigma') \} = \frac{2\pi \alpha'}{c} (\frac{1+\Gamma_{11}}{2}\CP_-)_{\alpha\beta} \delta ( \sigma-\sigma' )
\end{equation}
Hence
\begin{equation}
\{ \psi_{\alpha} (\tau,\sigma) , \psi_{\beta}^* (\tau,\sigma') \} = \frac{2\pi \alpha'}{c} (\frac{1+\Gamma_{11}}{2}( \CP_- + 2 \CP_+) )_{\alpha\beta} \delta ( \sigma-\sigma' )
\end{equation}

By introducing the fermionic momenta
\begin{equation}
\rho \equiv \frac{\delta \CL_2}{\delta \dot{\psi}} = - \frac{ic}{2} ( 2 \CP_- + \CP_+ ) \psi^*
\end{equation}
we can write the following anticommutation relation
\begin{equation}
\label{anticommu}
\{ \psi_{\alpha} (\tau,\sigma) , \rho_{\beta} (\tau,\sigma') \} = - 2\pi i \alpha' (\frac{1+\Gamma_{11}}{2} ( \CP_- + \CP_+) )_{\alpha\beta} \delta(\sigma-\sigma')
\end{equation}
The quadratic fermionic Hamiltonian is therefore
\begin{equation}
\label{CH2F}
\CH_{2,F} =
 \frac{i}{4 c^2} (c^2 \psi_+ \psi_+' -4 \rho_+ \rho_+' +2 c^2 \psi_- \psi_-' -2 \rho_- \rho_-')
 - \frac{i}{2} \psi_+ \rho_++i\psi_-\rho_-
 + \frac{1}{2} \psi_- \Gamma_{56} \rho_-
\end{equation}
where we have defined $\rho_{\pm}=\CP_{\pm}\rho$.
The fermionic spectrum can then be computed and reads
\begin{equation}
\label{fermppwave}
c H_{2,F} = \sum_{n\in \Z} \left[ \sum_{b=1}^4 \omega_n F_n^{(b)} +
\sum_{b=5}^6 \left( \Omega_n + \frac{c}{2} \right) F_n^{(b)} +
\sum_{b=7}^8 \left( \Omega_n - \frac{c}{2} \right) F_n^{(b)} \right]
\end{equation}
with the number operators $F^{(b)}_n= d_{n,\alpha}^\dagger d_{n,\alpha}$ for $b=1,\ldots ,4$, and
$F_n^{(b)} = b^\dagger_{n,\alpha} b_{n,\alpha}$ for $b=5,\ldots,8$. The level-matching condition, including also the bosonic part, is
\begin{equation}
\label{levelmbf} \sum_{n\in \Z}n \left[\sum_{i=1}^4
\hat{N}^i_n+\sum_{a=1}^2 \left(M_n^a + N_n^a\right) +\sum_{b=1}^8
F^{(b)}_n\right]
 = 0
\end{equation}
As mentioned before, the fermionic spectrum splits into four light and four heavy modes. The pp-wave spectrum \eqref{fermppwave} is in agreement with the one computed in \cite{Sugiyama:2002tf,Hyun:2002wu,Nishioka:2008gz} in a different coordinate system. However the result of~\cite{Sugiyama:2002tf,Hyun:2002wu,Nishioka:2008gz} has been obtained using the $\kappa$-symmetry gauge fixing condition $\Gamma^+\theta=0$. This differs from the gauge choice \eqref{fermgauge} used in this paper which is the appropriate one if one wants to study corrections to the pp-wave limit. In fact the gauge condition \eqref{fermgauge} ensures that we are selecting the appropriate supersymmetric states.

\section{Fermionic field redefinition on the Lagrangian}
\label{sec:lagr_redef}

The Lagrangian \eqref{CL2B}-\eqref{CL4B}, \eqref{CL2F}-\eqref{L4F1} found in Section \ref{sec:lclagr}
is the full Lagrangian to order $1/R^2$. However, it is convenient
to perform fermionic field redefinitions on this Lagrangian to make
it easier to pass to a Hamiltonian formalism. In particular, the
main complications for the fermionic terms in passing to the
Hamiltonian formalism are to change the fermionic variables to
fermionic positions and momenta and to perform the Dirac procedure.
For both these procedures the relevant quantities to consider are
the fermionic momenta. However, it is important to notice that it is
the fermionic momenta as functions of the bosonic positions and
momenta, as opposed to the bosonic positions and velocities, which
are the relevant quantities for both the change of fermionic
variables and the Dirac procedure.

The goal of performing fermionic field redefinitions of the
Lagrangian is thus that the fermionic momenta, with the bosonic
variables being the bosonic positions and momenta, are as simple as
possible in terms of their $1/R$ and $1/R^2$ corrections.
Unfortunately, it does not seem possible in a straightforward manner
to remove $1/R$ corrections to the fermionic momenta by fermionic
field redefinitions. This is because the field redefinitions induce
problematic terms with second derivatives of the bosonic coordinates
at order $1/R^2$. Therefore our goal in the following is to perform
field redefinitions of the Lagrangian such that the fermionic
momenta have no $1/R^2$ corrections.

Given the Lagrangian $\CL_{\rm gf} ( X^i , \dot{X}^i, {X'}^i ,
\theta , \dot{\theta}, \theta')$ consider now a field redefinition
$\tilde{\theta}= \tilde{\theta} ( X^i , \dot{X}^i, {X'}^i , \theta
)$ such that the new Lagrangian is given by
\begin{equation}
\CL_{\rm new} ( X^i , \dot{X}^i, {X'}^i , \theta, \dot{\theta} ,
\theta') = \CL_{\rm gf} ( X^i , \dot{X}^i, {X'}^i , \tilde{\theta} ,
\dot{\tilde{\theta}} , \tilde{\theta}')
\end{equation}
Take $\tilde{\theta}= \tilde{\theta} ( X^i , \dot{X}^i, {X'}^i ,
\theta )$  to be of the form
\begin{equation}
\label{fieldredef} \tilde{\theta} = \theta + \frac{1}{R^2} K^a
(\CP_+
 + \frac{1}{2} \CP_- ) \Gamma^0 \Gamma_a \theta + \frac{1}{R^2}
\tilde{K}^a (\CP_+ + \frac{1}{2} \CP_- ) \Gamma^0 \Gamma_{11}
\Gamma_a \theta
\end{equation}
where $a$ is summing over $+,-,1,2,...,8$ and $K^a=K^a( X^i ,
\dot{X}^i, {X'}^i , \theta )$, $\tilde{K}^a=\tilde{K}^a( X^i ,
\dot{X}^i, {X'}^i , \theta )$ do not carry any spinor indices. Then
we see that
\begin{equation}
\tilde{\theta} ( \CP_+ + 2 \CP_- ) \dot{\tilde{\theta}} = A_{+,\tau}
+ \frac{2}{R^2} ( K^a A_{a,\tau} +  \tilde{K}^a \tilde{A}_{a,\tau} )
+ \CO( R^{-3} )
\end{equation}
Notice that the terms with $\dot{K}^a$ and $\dot{\tilde{K}}^a$
vanish since $\bar{\theta} \Gamma_a \theta = 0$ and $\bar{\theta}
\Gamma_{11} \Gamma_a \theta = 0$. Similarly we have
\begin{equation}
\tilde{\theta} \Gamma^0 \Gamma_{11} \Gamma_+ \tilde{\theta}' =
\tilde{A}_{+,\sigma} + \frac{2}{R^2} ( K^a \tilde{A}_{a,\sigma} +
\tilde{K}^a A_{a,\sigma}
 ) + \CO( R^{-3} )
\end{equation}
\begin{equation}
\tilde{\theta} \Gamma^0 \Gamma_+ \Gamma_{56} \tilde{\theta} =
B_{+56} + \frac{2}{R^2} ( K^a B_{a56} + \tilde{K}^a \tilde{B}_{a56}
) + \CO (R^{-3})
\end{equation}
\begin{equation}
\tilde{\theta} \Gamma_{123} (  \CP_+ + 4 \CP_- ) \tilde{\theta} =
C_{++} +   \frac{2}{R^2} ( K^a C_{+a} - \tilde{K}^a \tilde{C}_{+a})  + \CO
(R^{-3})
\end{equation}
where we used Eqs.~\eqref{iden1}-\eqref{fourterms}. Thus, the
field redefinition \eqref{fieldredef} induces the following
additional terms that we should add to the original Lagrangian
\begin{equation}
\begin{array}{l} \ds
\frac{ic}{R^2} \Big[ K^a ( A_{a,\tau} + \tilde{A}_{a,\sigma} ) +
\tilde{K}^a ( \tilde{A}_{a,\tau}+ A_{a,\sigma} ) \Big] + \frac{ic^2}{4
R^2} \Big[ K^a (B_{a56}+B_{a78}) + \tilde{K}^a (\tilde{B}_{a56} + \tilde{B}_{a78}) \Big] \\[3mm] \ds +
\frac{ic^2}{2R^2} (- K^a C_{+a} + \tilde{K}^a \tilde{C}_{+a})
\end{array}
\end{equation}
We choose now the following field redefinition
\begin{equation}
j=1,2,3,4:\ \ K^j = - \frac{i}{2} C_{+j} + \frac{i}{2c}
\tilde{A}_{j,\sigma} \spa \tilde{K}^j = \frac{i}{2c} A_{j,\sigma}
\end{equation}
\begin{align}
j=5,6,7,8:\ \ & K^j = - \frac{1}{c} s_j u_4 \dot{X}^j - \frac{i}{2}
C_{+j} + \frac{i}{2c} \tilde{A}_{j,\sigma} - \frac{i}{8}
\sum_{k=5}^8 \epsilon_{jk} B_{k +-} + \frac{i}{4} s_j B_{4j+} \nn \\
& \tilde{K}^j = - \frac{1}{c} s_j u_4 {X'}^j + \frac{i}{2c}
A_{j,\sigma}
\end{align}
\begin{equation}
\begin{array}{c} \ds
K^+ = - \frac{1}{2c^2} \sum_{i=1}^8 \Big[ (\dot{X}^i)^2 + ({X'}^i)^2
\Big] - u_4^2 + \frac{i}{2c} \tilde{A}_{+,\sigma} - \frac{i}{4c}
\tilde{A}_{-,\sigma}   + \frac{i}{4} ( B_{+56} + B_{+78} - C_{++}
+  C_{+-}    ) \\[2mm] \ds \tilde{K}^+ = - \frac{1}{c^2} \sum_{i=1}^8 \dot{X}^i
{X'}^i - \frac{i}{2c} A_{+,\sigma} - \frac{i}{4c} A_{-,\sigma} \\[2mm]
\ds K^- = - \frac{1}{2c^2} \sum_{i=1}^8 \Big[ (\dot{X}^i)^2 +
({X'}^i)^2
\Big] + \frac{i}{4c} \tilde{A}_{+,\sigma} \\[2mm]
\ds \tilde{K}^- =  \frac{1}{c^2} \sum_{i=1}^8 \dot{X}^i {X'}^i +
\frac{i}{4c} A_{+,\sigma}
\end{array}
\end{equation}
This field redefinition gets rid of all $1/R^2$ terms in the
fermionic momenta when written in terms of the bosonic positions and
momenta, apart from terms coming from the $\bar{\theta} \Gamma_+
\CM^2 \dot{\theta}$ kinetic term in \eqref{L4F1}. To remove these
terms we perform the additional field redefinition
\begin{equation}
\tilde{\theta} = \theta - \frac{1}{12 R^2} \Big[ \bar{\theta} \Gamma_+ \CM^2 ( \CP_+ + \frac{1}{2} \CP_- ) \Big]^T
\end{equation}
This induces the following additional terms that we should add to
the four-fermion Lagrangian \eqref{L4F1}
\begin{equation}
- \frac{ic}{12 } \bar{\theta} \Gamma_+ \CM^2 \dot{\theta} + \frac{ic}{12 } \bar{\theta} \Gamma_+ \CM^2 \Gamma_{11} \theta' - \frac{ic^2}{48 } \bar{\theta} \Gamma_+ \CM^2 (\Gamma_{56} + \Gamma_{78} ) \theta + \frac{ic^2}{24 } \bar{\theta} \Gamma_+ \CM^2 P \Gamma_{0123} \Gamma_+ \theta
\end{equation}
We now get the following modified Lagrangian after the field
redefinition
\begin{equation}
\label{newCL4BF}
\begin{array}{l} \ds
\CL_{4,BF} = -  \frac{i}{4} \sum_{i=1}^8 \Big[ (\dot{X}^i)^2 + ({X'}^i)^2 \Big] \Big( \frac{4}{c} \tilde{A}_{+,\sigma}  + B_{+56} + B_{+78}  - C_{++} + C_{+-} \Big)
- \frac{ic^2}{2} \sum_{i=1}^3 u_i^2 C_{++}
\\[2mm] \ds
+ ic \tilde{A}_{+,\sigma}
\Big[
\sum_{i=1}^3 u_i^2 -u_4^2 \Big]
+ \frac{ic^2}{4} \sum_{i=1}^4  u_i^2  ( B_{+56} + B_{+78} )
- 2 i u_4 \sum_{i=5}^8 s_i ( A_{i,\tau} +
\tilde{A}_{i,\sigma} ) \dot{X}^i
\\[2mm] \ds
- \frac{i}{2} \sum_{i=1}^8 \dot{X}^i {X'}^i \Big( \frac{4}{c} A_{+,\sigma} + \tilde{B}_{+56}+\tilde{B}_{+78}  + \tilde{C}_{+-} \Big)
+ i \sum_{i,j=1}^8 \Big[  C_{ij} ({X'}^i {X'}^j - \dot{X}^i \dot{X}^j ) + 2  \tilde{C}_{ij} {X'}^i \dot{X}^j \Big]
\\[2mm] \ds
+ \frac{i c}{2} u_4 \sum_{i=5}^8 s_i ( 3 C_{+i} \dot{X}^i + \tilde{C}_{+i} {X'}^i )
+ \frac{ic}{2} u_4 \sum_{i=1}^8 \Big[ (B_{i56} - B_{i78}) \dot{X}^i - ( \tilde{B}_{i56}  - \tilde{B}_{i78}){X'}^i \Big]
\\[2mm] \ds
- \frac{ic}{2} \sum_{i=1}^8 \sum_{j=1}^4 u_j ( B_{-ij} \dot{X}^i - \tilde{B}_{-ij} {X'}^i )
- \frac{i}{2} \sum_{i=1}^8 \sum_{j=5}^8 s_j \Big[ ( \dot{X}^i \dot{X}^j - {X'}^i {X'}^j ) B_{4 i j} + ( \dot{X}^i {X'}^j - {X'}^i \dot{X}^j ) \tilde{B}_{4 i j} \Big]
\\[2mm] \ds
+ \frac{ic}{2} \sum_{i=1}^3 \sum_{j=4}^8 u_i \Big[ B_{+ij} \dot{X}^j -  \tilde{B}_{+ij} {X'}^j \Big]
- \frac{ic}{4} ( B_{+56} \dot{x}_1 y_1 +  \tilde{B}_{+56} x_1' y_1 +  B_{+78} \dot{x}_2 y_2 +  \tilde{B}_{+78} x_2' y_2 )
\\[2mm] \ds
+\frac{i}{4} \sum_{i=1}^8 \sum_{j,k=5}^8 \epsilon_{jk} \Big[ (B_{+ij} - B_{-ij}) ( \dot{X}^i \dot{X}^k - {X'}^i {X'}^k )  + (\tilde{B}_{+ij}-\tilde{B}_{-ij}) ( \dot{X}^i {X'}^k - {X'}^i \dot{X}^k )  \Big]
\\[2mm] \ds
+ \frac{ic}{4} u_4 \sum_{i=5}^8 (- B_{+4i} \dot{X}^i + 3 \tilde{B}_{+4i} {X'}^i ) %
- \frac{ic}{8} u_4 \sum_{i,j=5}^8 s_i \epsilon_{ij} ( B_{+i-} \dot{X}^j + \tilde{B}_{+i-} {X'}^j )
\\[2mm] \ds
+ \frac{ic}{2} u_4 \sum_{i=1}^3 ( B_{+i4} \dot{u}_i - \tilde{B}_{+i4} u_i' )
+ i c \sum_{i,j=1}^3 u_i' u_j \tilde{B}_{+ij}
- \frac{ic}{2} u_4 \sum_{i=5}^8 s_i
(B_{i56}+B_{i78}) \dot{X}^i
\end{array}
\end{equation}
\begin{equation}
\label{newCL4F}
\begin{array}{l} \ds
\CL_{4,F} =  \frac{ic}{12} \bar{\theta} \Gamma_{11} \Gamma_+ \CM^2
\theta' + \frac{ic}{12 } \bar{\theta} \Gamma_+ \CM^2 \Gamma_{11}
\theta'
- \frac{1}{2} \sum_{i=1}^8 \Big[ A_{i,\tau} + \tilde{A}_{i,\sigma} +
\frac{c}{4} ( B_{i56}+B_{i78} ) - c C_{+i} \Big]^2
\\[2mm] \ds
+ \frac{c^2}{8} \sum_{i=1}^8 C_{+i}^2
- \frac{c}{4} \sum_{i=5}^8 \Big[ s_i B_{+4i} - \frac{1}{2}
\sum_{j=5}^8 \epsilon_{ij} B_{+-j} \Big] \Big( A_{i,\tau} +
\tilde{A}_{i,\sigma} - \frac{c}{2} C_{+i} + \frac{c}{4} (
B_{i56}+B_{i78} ) \Big)
\\[2mm] \ds
+ \frac{1}{2} ( A_{+,\sigma}^2 - \tilde{A}_{+,\sigma} ^2 )
+ \frac{c}{4}
A_{+,\sigma} \Big[   \tilde{C}_{+-} + \tilde{B}_{+56}+\tilde{B}_{+78}  \Big]
-\frac{c}{4} \tilde{A}_{+,\sigma} \Big[  C_{+-} - C_{++} +  B_{+56} + B_{+78}
 \Big]
\end{array}
\end{equation}

In terms of this new Lagrangian we have
\begin{equation}
\label{simplefermmom}
\begin{array}{c} \ds
\frac{\partial \CL}{\partial A_{j,\tau}} = \frac{i}{R} p_j \spa \frac{\partial \CL}{\partial \tilde{A}_{j,\tau}} = - \frac{i}{R} {X'}^j \\[5mm] \ds \frac{\partial \CL}{\partial A_{+,\tau}} = \frac{ic}{2} \spa \frac{\partial \CL}{\partial \tilde{A}_{+,\tau}} = \frac{\partial \CL}{\partial A_{-,\tau}} = \frac{\partial \CL}{\partial \tilde{A}_{-,\tau}} = 0 \\[5mm] \ds
\frac{\partial \CL}{\partial ( \bar{\theta} \Gamma_+ \CM^2 \dot{\theta}) } = 0
\end{array}
\end{equation}
with $j=1,2,...,8$ and where we used
\begin{equation}
p_{i=1...4} = (p_{u_1},p_{u_2},p_{u_3},p_{u_4}) \spa p_{i=5...8}
= \frac{\sqrt{2}}{4} (p_{x_1},p_{y_1},p_{x_2},p_{y_2})
\end{equation}
From \eqref{simplefermmom} we see that the purpose of making the fermionic field redefinitions is fulfilled in that the fermionic momenta do not have $1/R^2$ corrections and that they do not depend on $\dot{\theta}$. This ensures that one needs only to take into account the $1/R$ corrections to the fermionic momenta when changing variables to fermionic phase space variables in the Hamiltonian and when performing the Dirac procedure. Furthermore any $\dot{\theta}$ dependence in the fermionic momenta would have caused problems in these procedures.

\section{Full light cone Hamiltonian up to quartic terms}
\label{sec:lcham}

This section contains the two principle results of this paper. The first is a way to combine the shift to fermionic phase space variables and the Dirac procedure (getting the Hamiltonian in a form where the canonical quantization is simple). This is a highly non-trivial result since we have both a first and a second order correction to the Hamiltonian.

The second principle result is that we have computed explicitly the full Hamiltonian, including quadratic, cubic and quartic terms in the fields, in phase space variables with a simple canonical quantization procedure.
The total Hamiltonian that we compute in this paper is written in the following way
\begin{equation}
\CH = \CH_{2,B} + \CH_{2,F} + \frac{1}{R} (\CH_{3,B} + \CH_{3,BF}) +
\frac{1}{R^2} (\CH_{4,B} + \CH_{4,BF} + \CH_{4,F}) + \CO ( R^{-3} )
\end{equation}
%

\subsection{Qubic and quartic terms in bosonic Hamiltonian}

The bosonic terms in the Hamiltonian are readily obtained using the Legendre transform
\begin{equation}
c \CH_B = \sum_{i=1}^4 \dot{u}_i \Pi_{u_i} + \sum_{i=1}^2 \dot{x}_i \Pi_{x_i} + \sum_{i=1}^2 \dot{y}_i \Pi_{y_i} - \CL_B
\end{equation}
where $\CL_B$ is the bosonic part of the light cone gauge fixed Lagrangian \eqref{CL2B}-\eqref{CL4B} and where
\begin{equation}
\Pi_\mu = \frac{\partial \CL_B}{\partial \dot{x}^\mu}
\end{equation}
For convenience we define as in Section \ref{sec:ppwave}
\begin{equation}
p_{x_a}  \equiv 8 \Pi_{x_a} - c y_a \spa p_{y_a} \equiv 8 \Pi_{y_a}
\spa p_{u_i}  \equiv \Pi_{u_i}
\end{equation}
Notice that these fields are functions of the momenta and position
variables. The cubic terms in the bosonic Hamiltonian are found as
\begin{equation}
\label{CH3B}
\CH_{3,B} = \frac{u_4}{8c}  \Big[ p_{x_1}^2 + p_{y_1}^2 - p_{x_2}^2
- p_{y_2}^2  - {x_1'}^2 - {y_1'}^2 + {x_2'}^2 + {y_2'}^2 \Big]
\end{equation}
The quartic terms in the bosonic Hamiltonian are
\begin{eqnarray}
\label{CH4B} && \CH_{4,B} = \frac{2}{c^3}(\sum_{i=1}^8 p_i
{X'}^i)^2-\frac{1}{2 c^3}\left(\sum_{i=1}^8( p_i^2+({X'}^i)^2)-c^2
\sum_{i=1}^3 u_i^2 + c^2u_4^2\right)^2 \cr && + c (\sum_{i=1}^3
u_i^2)^2+\frac{4}{3}cu_4^4+\frac{1}{c}\sum_{i,j=1}^3
u_i^2({u_j'}^2-p_j^2)+\frac{2}{c}u_4^2 \sum_{i=5}^8 p_i^2 \cr
&&+\frac{1}{12\sqrt{2}}(p_5y_1^3+p_7 y_2^3)+\frac{1}{2c}y_1^2
(p_5^2-{X_5'}^2)+\frac{1}{2c}y_2^2 (p_7^2-{X_7'}^2)
\end{eqnarray}
with
\begin{equation}
\begin{array}{c}
\ds p_{i=1...4} = (p_{u_1},p_{u_2},p_{u_3},p_{u_4}) \spa p_{i=5...8}
= \frac{\sqrt{2}}{4} (p_{x_1},p_{y_1},p_{x_2},p_{y_2}) \\
\ds {X'}^{i=1...4} = (u_1',u_2',u_3',u_4') \spa {X'}^{i=5...8} =
\frac{\sqrt{2}}{4} ( x_1',y_1',x_2',y_2' )
\end{array}
\end{equation}
Thus, \eqref{CH2B} along with \eqref{CH3B}-\eqref{CH4B} constitute the final expression for the bosonic part of the Hamiltonian.

\subsection{Preliminary expressions for the fermionic Hamiltonian}
\label{sec:premham}

In this section we compute the fermionic part of the Hamiltonian as function of the bosonic phase space variables ($i.e.$ position and momenta) but without changing variables to fermionic phase space variables. This change of variables is performed below in Section \ref{sec:hamshift} where we also perform the Dirac procedure. We find the Hamiltonian using the field redefined Lagrangian given by \eqref{finalCL2F}, \eqref{CL3BF} and \eqref{newCL4BF}-\eqref{newCL4F}.

Before finding the Hamiltonian we perform first the variable shift in the Lagrangian from $\theta$ to $\psi$ and $\psi^*$ defined by
\begin{equation}
\label{defpsi2}
\psi = \theta^1 + i \Gamma_{049} \theta^2 \spa \psi^* = \theta^1 - i \Gamma_{049} \theta^2
\end{equation}
as in Eq.~\eqref{defpsi} in Section \ref{sec:ppwave}. In terms of the Lagrangian in these variables we define the fermionic momenta
\begin{equation}
\label{defrhos}
\rho = \frac{\partial \CL_{\rm gf}}{\partial \dot{\psi}} \spa
\rho^* = \frac{\partial \CL_{\rm gf}}{\partial \dot{\psi}^*}
\end{equation}
We find $\rho$ and $\rho^*$ explicitly in Section \ref{sec:hamshift}.
The Hamiltonian is now found by employing the Legendre transform
\begin{equation}
\label{legform}
c \CH = \dot{\psi} \rho + \dot{\psi}^* \rho^* +   \sum_{i=1}^4 \dot{u}_i \Pi_{u_i} + \sum_{i=1}^2 \dot{x}_i \Pi_{x_i} + \sum_{i=1}^2 \dot{y}_i \Pi_{y_i} - \CL_{\rm gf}
\end{equation}
We can perform the Legendre transform without knowing explicitly $\rho$ and $\rho^*$ in terms of $\psi$ and $\psi^*$.
All we need to employ is that we know from \eqref{simplefermmom} that when the Lagrangian is written in bosonic phase space variables there are no terms that are non-linear in $\dot{\theta}$. Hence the fermionic part in the Legendre transform \eqref{legform} simply corresponds to removing all terms with $\dot{\theta}$ in the Lagrangian written in bosonic phase space variables.

From the above we compute the cubic fermionic part of the Hamiltonian to be%
\footnote{In deriving this we used that $B_{+-i}=0$ for $i=1,2,3$.}
\begin{equation}
\label{pH3BF}
\begin{array}{l} \ds
\CH_{3,BF} =
 - \frac{ic}{4}  B_{+-4} u_4 - \frac{ic}{4}
( B_{+56} - B_{+78} ) u_4    - \frac{i }{4} \sum_{i=5}^8 s_i ( B_{+4i} p_i +
\tilde{B}_{+4i} {X'}^i ) \\[2mm] \ds + i \sum_{i=1}^8 \Big[ \Big( - \frac{1}{c} \tilde{A}_{i,\sigma} + C_{+i} - \frac{1}{4} ( B_{i56}+B_{i78}) \Big) p_i + \Big( \frac{1}{c} A_{i,\sigma} + \tilde{C}_{+i} + \frac{1}{4} ( \tilde{B}_{i56}+\tilde{B}_{i78}) \Big) {X'}^i \Big]
\\[2mm] \ds
- \frac{i}{8} \sum_{i,j=5}^8 \epsilon_{ij} ( B_{+-i} p_j + \tilde{B}_{+-i} {X'}^j )
\end{array}
\end{equation}
The part with two bosons and two fermions is
\begin{equation}
\label{pH4BF}
\begin{array}{l} \ds
\CH_{4,BF} = \frac{i}{c^2}
\sum_{i=1}^8 \Big( p_i^2 + ({X'}^i)^2 \Big) \Big[ \tilde{A}_{+,\sigma}
 + \frac{c}{4} ( B_{+56} + B_{+78}  -  C_{++}  +  C_{+-} ) \Big]
- i \tilde{A}_{+,\sigma} \Big[ \sum_{i=1}^3 u_i^2 -u_4^2 \Big]
\\[2mm] \ds
+ \frac{2i}{c^2} \sum_{i=1}^8 p_i  {X'}^i \Big[  A_{+,\sigma} +
\frac{c}{4} ( \tilde{B}_{+56}+\tilde{B}_{+78}) + \frac{c}{4}
\tilde{C}_{+-}   \Big]
+ \frac{ic}{2} \sum_{i=1}^3 u_i^2 C_{++}
- \frac{ic}{4} \sum_{i=1}^4 u_i^2 ( B_{+56} + B_{+78})
\\[2mm] \ds
+ \frac{i}{2} u_4 \sum_{i=5}^8 s_i \Big[ C_{+i}  p_i  -
\tilde{C}_{+i} {X'}^i \Big]
- \frac{i}{c} \sum_{i,j=1}^8 \Big[  C_{ij} ({X'}^i {X'}^j - p_i  p_j
) + 2  \tilde{C}_{ij} {X'}^i p_j  \Big]
- i \sum_{i,j=1}^3 u_i' u_j \tilde{B}_{+ij}
\\[2mm] \ds
- \frac{i}{2} u_4 \sum_{i=1}^8 \Big[ (B_{i56} - B_{i78}) p_i  - (
\tilde{B}_{i56}  - \tilde{B}_{i78}){X'}^i \Big]
- \frac{i}{8} u_4 \sum_{i,j=5}^8 s_i \epsilon_{ij} ( 3 B_{+-i} p_j
+ \tilde{B}_{+-i} {X'}^j )
\\[2mm] \ds
- \frac{i}{2} \sum_{i=1}^4 \sum_{j=1}^8 u_i \Big[ B_{-ij} p_j  - \tilde{B}_{-ij} {X'}^j \Big]
+ \frac{i}{2c} \sum_{i=1}^8 \sum_{j=5}^8 s_j \Big[ ( p_i  p_j  - {X'}^i {X'}^j ) B_{4 i j} + ( p_i  {X'}^j - {X'}^i p_j  ) \tilde{B}_{4 i j} \Big]
\\[2mm] \ds
- \frac{i}{2} \sum_{i=1}^3 \sum_{j=4}^8 u_i \Big[ B_{+ij} p_j -  \tilde{B}_{+ij} {X'}^j \Big]
+ \frac{i}{4} ( B_{+56} p_{x_1} y_1 +  \tilde{B}_{+56} x_1' y_1 +  B_{+78} p_{x_2} y_2 +  \tilde{B}_{+78} x_2' y_2 )
\\[2mm] \ds
- \frac{i}{4c} \sum_{i=1}^8 \sum_{j,k=5}^8 \epsilon_{jk} \Big[
(B_{+ij} - B_{-ij}) ( p_i  p_k  - {X'}^i {X'}^k )  +
(\tilde{B}_{+ij}-\tilde{B}_{-ij}) ( p_i  {X'}^k - {X'}^i p_k  )
\Big]
\\[2mm] \ds
- \frac{i}{4} u_4 \sum_{i=5}^8 ( B_{+4i} p_i  + 3 \tilde{B}_{+4i} {X'}^i )
+ \frac{i}{2} u_4 \sum_{i=1}^3 ( B_{+4i} p_i - \tilde{B}_{+4i} u_i' )
\end{array}
\end{equation}
Finally, the four-fermion part is
\begin{equation}
\label{pH4F}
\begin{array}{l} \ds
\CH_{4,F} = - \frac{i}{12} \Big( \bar{\theta} \Gamma_{11} \Gamma_+
\CM^2 \theta' + \bar{\theta} \Gamma_+ \CM^2 \Gamma_{11} \theta'
\Big)
- \frac{1}{2c} ( A_{+,\sigma}^2 - \tilde{A}_{+,\sigma}^2 )
\\[2mm] \ds
- \frac{1}{4} A_{+,\sigma} ( \tilde{C}_{+-} + \tilde{B}_{+56} + \tilde{B}_{+78} )
+ \frac{1}{4} \tilde{A}_{+,\sigma} ( C_{+-} - C_{++} + B_{+56} + B_{+78} )
\\[2mm] \ds
- \frac{c}{8} \sum_{i=1}^4 C_{+i}^2
- \frac{c}{32} \sum_{i=5}^8 \Big[ 2 C_{+i} - s_i B_{+4i} + \frac{1}{2} \sum_{j=5}^8 \epsilon_{ij} B_{+-j} \Big]^2
\end{array}
\end{equation}
Note that all the above expressions \eqref{pH3BF}-\eqref{pH4F} can be thought of as being in terms of $\psi$ and $\psi^*$ rather than in terms of $\theta$. More explicitly, one can invert \eqref{defpsi2}
\begin{equation}
\label{inverttheta}
\theta = \theta^1 + \theta^2 \spa \theta^1 = \frac{1}{2} ( \psi+\psi^* ) \spa \theta^2 = \frac{\Gamma_{049}}{2i} ( \psi-\psi^* )
\end{equation}
and plug this into \eqref{ABCdef} in order to obtain $A_{a,\sigma}$, $\tilde{A}_{a,\sigma}$, $B_{abc}$, $\tilde{B}_{abc}$, $C_{ab}$ and $\tilde{C}_{ab}$ in terms of $\psi$ and $\psi^*$.

\subsection{Dirac procedure and shift to fermionic phase space variables}
\label{sec:hamshift}

As stated in Section \ref{sec:premham} the Hamiltonian \eqref{pH3BF}-\eqref{pH4F} is not our final expression for the fermionic terms in the Hamiltonian. To find the final form we need to perform two tasks, both involving a fermionic field redefinition of the Hamiltonian. First we should make the coordinate change from $(\psi,\psi^*)$ to the fermionic phase space variables $(\psi,\rho)$. However, this is not sufficient since the canonical commutation relations would have a complicated structure due to the fact that the Dirac brackets receive $1/R$ and $1/R^2$ corrections. Hence the second task is to determine the field redefinition that one should perform on the Hamiltonian in order to be able to use the canonical commutation relations without $1/R$ and $1/R^2$ corrections.

Performing these two tasks, and finding the two fermionic field redefinitions, is very non-trivial and involved since we have both $1/R$ and $1/R^2$ corrections. We found therefore a way to combine the two field redefinitions into one field redefinition performed on the $\theta$ variable. In this way the structure of the combined field redefinition becomes sufficiently elegant and simple such that we can perform it explicitly and compute the resulting extra terms that one should add to the Hamiltonian \eqref{pH3BF}-\eqref{pH4F}.

\subsubsection*{Computing the fermionic momenta}

To proceed with the program outlined above we need to compute the fermionic momenta as defined in \eqref{defrhos}. From the field redefined Lagrangian \eqref{finalCL2F}, \eqref{CL3BF} and \eqref{newCL4BF}-\eqref{newCL4F} we see that the terms with $\dot{\theta}$ are
\begin{equation}
\CL_{\rm gf} =-  \frac{ic}{2} \dot{\psi} ( \CP_+ + 2\CP_-)\psi^*  +  \frac{i}{R} \sum_{i=1}^8 ( p_i A_{i,\tau} - {X'}^i \tilde{A}_{i,\tau} ) + \CO( R^{-3} )
\end{equation}
From this we find the fermionic momenta to be of the form
\begin{equation}
\label{fermmomenta1}
\rho = E ( \psi^* + m \psi^* + n \psi ) \spa \rho^* = E ( m \psi + n \psi^* )
\end{equation}
with $E$, $m$ and $n$ determined as
\begin{equation}
\label{fermmomenta2}
\begin{array}{c}\ds
E = - \frac{ic}{2} ( \CP_+ + 2\CP_- )
\\[2mm] \ds
m+n = \frac{1}{cR} \sum_{i=1}^8 ( p_i + {X'}^i ) ( \CP_+ + \frac{1}{2} \CP_- ) \Gamma^0 \Gamma_i ( \CP_+ + \CP_- )
\\[2mm] \ds\Gamma_{049}(m-n)\Gamma_{049} = \frac{1}{cR} \sum_{i=1}^8 ( p_i - {X'}^i ) ( \CP_+ + \frac{1}{2} \CP_- ) \Gamma^0 \Gamma_i ( \CP_+ + \CP_- )
\end{array}
\end{equation}

\subsubsection*{Shift to fermionic phase space variables}

Consider the Hamiltonian listed in Eqs.~\eqref{pH3BF}-\eqref{pH4F}. This is the fermionic part of the Hamiltonian written in terms of $\psi$ and $\psi^*$ as well as in the bosonic phase space variables. We write this as
\begin{equation}
\CH_{(1)} ( \psi, \psi^* )
\end{equation}
Our first task is to write this Hamiltonian in the phase space variables $\psi$ and $\rho$. This one can do by eliminating $\psi^*$ order by order in $1/R$ using the expressions for $\rho(\psi,\psi^*)$ in \eqref{fermmomenta1}-\eqref{fermmomenta2}. This can equivalently be thought of as a fermionic field redefinition of $\psi$ and $\psi^*$. Specifically, by inverting \eqref{fermmomenta1} the field redefinition takes the form
\begin{equation}
\label{shiftredef}
\psi (\psi,\rho) = \psi \spa \psi^* (\psi,\rho) = E^{-1} \rho - m E^{-1} \rho - n \psi + m^2 E^{-1} \rho + mn \psi
\end{equation}
Here $E^{-1}$ is defined by $E E^{-1} = E^{-1} E = \CP_+ + \CP_-$.
Then we define the Hamiltonian
\begin{equation}
\CH_{(2)} ( \psi, \rho ) \equiv  \CH_{(1)} ( \psi , \psi^* (\psi,\rho) )
\end{equation}
This is the Hamiltonian written correctly in terms of the phase space variables $\psi$ and $\rho$.

\subsubsection*{Dirac procedure}

The second task is to implement the Dirac procedure. This is necessary since the fermionic momenta \eqref{fermmomenta1}-\eqref{fermmomenta2} results in a complicated structure for the canonical commutation relations since the Dirac brackets are non-trivial.

We begin by defining the Poisson bracket for Grassmanian fields as
\begin{equation}
\{ A, B \}_P = - ( \CP_- + \CP_+ )_{\alpha\beta} \left[ \left( \frac{\partial A}{\partial \psi_\alpha }
\frac{\partial B}{\partial \rho_\beta } + \frac{\partial
B}{\partial \psi_\alpha } \frac{\partial A}{\partial \rho_\beta }
\right) + \left( \frac{\partial A}{\partial \psi^\dagger_\alpha }
\frac{\partial B}{\partial \rho^\dagger_\beta } + \frac{\partial
B}{\partial \psi^\dagger_\alpha } \frac{\partial A}{\partial
\rho^\dagger_\beta } \right) \right]
\end{equation}
Here the projector $\CP_++\CP_-$ is due to the fact that we work in the subspace of physical states.
From the fermionic momenta \eqref{fermmomenta1}-\eqref{fermmomenta2} we read off the constraints (up to corrections of order $R^{-3}$)
\begin{equation}
\begin{array}{l} \ds
\eta^1 (\psi,\psi^*,\rho,\rho^*) = \rho - E \psi^* -  E m \psi^* - E n \psi \\[2mm] \ds \eta^2 (\psi,\psi^*,\rho,\rho^*) = \rho^* - E m \psi - E n  \psi^* \end{array}
\end{equation}
In terms of these constraints the Dirac bracket for Grassmanian fields is defined as
\begin{equation}
\{ A,B \}_D = \{ A,B \}_P - \{ A, \eta^i_\alpha \}_P (C^{-1})^{(i\alpha)(j\beta)} \{ \eta^j_\beta , B\}_P
\end{equation}
where $C_{(i\alpha)(j\beta)} = \{ \eta^i_\alpha, \eta^j_\beta \}_P$. We compute
\begin{equation}
C = \matrto{0}{E}{E}{0} +  2 E \matrto{n }{m}{ m}{n}
\end{equation}
The inverse is
\begin{equation}
C^{-1} = \matrto{0}{E^{-1}}{E^{-1}}{0} -  2 \matrto{n}{m}{m}{n} E^{-1} + 4 \matrto{mn + nm }{m^2 + n^2}{m^2 + n^2}{mn + nm } E^{-1}
\end{equation}
up to corrections of order $R^{-3}$.
Using these results we find the Dirac brackets
\begin{equation}
\label{dbr}
\begin{array}{c} \ds
\{ \psi_\alpha , \psi_\beta \}_D = \big[ 2 n E^{-1} - 4 (mn+nm) E^{-1} \big]_{\alpha\beta}
\\[3mm] \ds
\{ \psi_\alpha , \rho_\beta \}_D = \{ \psi_\alpha , \rho_\beta \}_P + \big[ m  - 2 m^2 - 2n^2
\big]_{\alpha\beta}
\end{array}
\end{equation}
To quantize the theory, one should impose the above Dirac brackets
for $\psi$ and $\rho$ as the anti-commutators for the quantum
fields. However, one can also find a field redefinition
$\tilde{\psi} ( \psi , \rho )$ and $\tilde{\rho} ( \psi , \rho )$
such that the Poisson brackets for $\tilde{\psi}$ and $\tilde{\rho}$
are equal to the Dirac brackets \eqref{dbr}, $i.e.$
\begin{equation}
\label{DBcon}
\{ \psi_\alpha , \psi_\beta \}_D = \{
\tilde{\psi}_\alpha , \tilde{\psi}_\beta \}_P \spa \{ \psi_\alpha , \rho_\beta \}_D = \{ \tilde{\psi}_\alpha ,
\tilde{\rho}_\beta \}_P
\end{equation}
We write this field redefinition as
\begin{equation}
\label{diracredef}
\tilde{\psi} ( \psi , \rho ) = \psi + A \psi + B \rho \spa \tilde{\rho}  (\psi,\rho) = \rho + \tilde{A} \rho + \tilde{B} \psi
\end{equation}
and define the Hamiltonian
\begin{equation}
\label{hamdiracredef}
\CH_{(3)} ( \psi,\rho ) \equiv \CH_{(2)} ( \tilde{\psi} ( \psi , \rho ), \tilde{\rho} (\psi,\rho) )
\end{equation}
Note here that $\CH_{(2)}(\psi,\rho)$ obviously is the Hamiltonian that one should perform the field redefinition in since this is the one written in terms of the phase space variables.
We compute
\begin{eqnarray}
\label{psistartransf}
\psi^* ( \tilde{\psi} ( \psi , \rho ), \tilde{\rho} (\psi,\rho))
&=& E^{-1} \rho + ( - m E^{-1} + E^{-1} \tilde{A} - m E^{-1} \tilde{A} - n B + m^2 E^{-1} ) \rho \nn \\ && + ( - n +  E^{-1} \tilde{B} - n A  - m E^{-1} \tilde{B} + mn ) \psi
\end{eqnarray}
Imposing now \eqref{DBcon} we see that the field redefinition \eqref{diracredef}
 is required to satisfy
\begin{equation}
\begin{array}{c} \ds
\{ \tilde{\psi}_\alpha, \tilde{\psi}_\beta \}_P = \{ \psi_\alpha,\psi_\beta \}_D = [ (2n-4mn-4nm) E^{-1} ]_{\alpha\beta} \\[3mm] \ds \{ \tilde{\psi}_\alpha , \tilde{\rho}_\beta \}_P = \{ \psi_\alpha,\rho_\beta \}_D = \{ \psi_\alpha,\rho_\beta \}_P + [ m - 2m^2 - 2n^2 ]_{\alpha\beta}
\end{array}
\end{equation}
We compute
\begin{equation}
\begin{array}{c} \ds
\{ \tilde{\psi}_\alpha, \tilde{\psi}_\beta \}_P = - [ B + B^T + A B^T + B A^T ]_{\alpha\beta}
\\[2mm] \ds
\{ \tilde{\psi}_\alpha , \tilde{\rho}_\beta \}_P = \{ \psi_\alpha,\rho_\beta \}_P - [ A + \tilde{A}^T +  A \tilde{A}^T + B \tilde{B}^T ]_{\alpha\beta}
\end{array}
\end{equation}
and hence we obtain explicitly the constraints on the field redefinition \eqref{diracredef}
\begin{equation}
\label{matreqs}
\begin{array}{c} \ds
B + B^T + A B^T + B A^T = ( -2n + 4mn + 4nm) E^{-1}
\\[3mm] \ds
A + \tilde{A}^T +  A \tilde{A}^T + B \tilde{B}^T = - m + 2m^2 + 2n^2
\end{array}
\end{equation}

\subsubsection*{Combining change of variables and Dirac procedure into redefinition of $\theta$}

We consider now how to solve the constraints \eqref{matreqs} on the field redefinition \eqref{diracredef} in order for \eqref{DBcon} to be satisfied.
We do this by seeing the two field redefinitions \eqref{shiftredef} and \eqref{diracredef} as one combined redefinition from $\CH_{(1)}$ to $\CH_{(3)}$
\begin{equation}
\label{twofieldredef}
\CH_{(3)} ( \psi,\rho ) \equiv \CH_{(2)} ( \tilde{\psi} ( \psi , \rho ), \tilde{\rho} (\psi,\rho) ) = \CH_{(1)} ( \tilde{\psi} ( \psi , \rho ) , \psi^* ( \tilde{\psi} ( \psi , \rho ), \tilde{\rho} (\psi,\rho) )
\end{equation}
and demand that this field redefinition should take a simple and elegant form when written as a field redefinition of the variable $\theta$.

We begin with imposing that to first order in $1/R$ our transformation looks like
\begin{equation}
\tilde{\psi} (\psi,\rho) = \psi - m \psi - n E^{-1} \rho + \CO(R^{-2}) \spa
\psi^* ( \tilde{\psi} ( \psi , \rho ), \tilde{\rho} (\psi,\rho)) = E^{-1} \rho - m E^{-1} \rho - n \psi + \CO(R^{-2})
\end{equation}
From \eqref{psistartransf} we see that this means that $\tilde{A}$ and $\tilde{B}$ are zero at order $1/R$. We get furthermore that
\begin{equation}
A = - m + \CO(R^{-2} ) \spa B = - n E^{-1} + \CO(R^{-2} )
\end{equation}
using here $(nE^{-1})^T = n E^{-1}$.
Inserting this in \eqref{matreqs} we get
\begin{equation}
A  = - m + 2 m^2 + 2 n^2 - \tilde{A}^T \spa B E = - n + \frac{3}{2} ( m n + n m )
\end{equation}
using here $n E^{-1} m^T = nm E^{-1}$.
We get therefore
\begin{equation}
\tilde{\psi} (\psi,\rho) = \psi - m \psi - n E^{-1} \rho + (2m^2+2n^2-\tilde{A}^T) \psi + \frac{3}{2} (mn+nm) E^{-1} \rho
\end{equation}
\begin{eqnarray} \psi^* ( \tilde{\psi} ( \psi , \rho ), \tilde{\rho} (\psi,\rho))
&=& E^{-1} \rho - m E^{-1} \rho - n \psi + (m^2+n^2 +  E^{-1} \tilde{A} E)E^{-1} \rho \nn \\ && + (   n m  + mn + E^{-1} \tilde{B} ) \psi
\end{eqnarray}

We write now the combined field redefinition of the Hamiltonian \eqref{twofieldredef} as
\begin{equation}
\CH_{(3)} ( \psi , E \psi^* ) = \CH_{(1)} ( \psi + \Delta \psi (\psi,\psi^*) , \psi^* + \Delta \psi^* (\psi,\psi^*) )
\end{equation}
Then
\begin{equation}
\begin{array}{c} \ds
\Delta \psi (\psi,\psi^*) = - m \psi - n \psi^* + (2m^2+2n^2-\tilde{A}) \psi + \frac{3}{2} (mn+nm) \psi^*
\\[3mm] \ds
\Delta \psi^* (\psi,\psi^*) = - m \psi^* - n \psi + (m^2+n^2 +  E^{-1} \tilde{A} E) \psi^* + (   n m  + mn + E^{-1} \tilde{B} ) \psi
\end{array}
\end{equation}
We would like that $\Delta \psi^* (\psi,\psi^*) = \Delta \psi ( \psi^* , \psi )$ since this seems a necessary requirement for writing the field redefinition in a simple fashion in terms of $\theta$. This fixes
\begin{equation}
\tilde{A}^T = E^{-1} \tilde{A} E = \frac{1}{2} ( m^2+n^2 ) \spa E^{-1} \tilde{B} = \frac{1}{2} ( mn+ nm )
\end{equation}
Note that this works since $E^{-1} (m^2)^T E = m^2$ and $E^{-1} (n^2)^T E = n^2$.
Then the total field redefinition is
\begin{equation}
\label{totredef}
\begin{array}{c} \ds
\Delta \psi (\psi,\psi^*) = - m \psi - n \psi^* + \frac{3}{2} (m^2+n^2) \psi + \frac{3}{2} (mn+nm) \psi^*
\\[3mm] \ds
\Delta \psi^* (\psi,\psi^*) = - m \psi^* - n \psi + \frac{3}{2} (m^2+n^2) \psi^* + \frac{3}{2} (mn+nm) \psi
\end{array}
\end{equation}
We now wish to write down the equivalent field redefinition in terms of $\theta$. Thus, we wish to perform the field redefinition
\begin{equation}
\CH_{(3)} ( \theta ) = \CH_{(1)} ( \theta + \Delta \theta ( \theta ) )
\end{equation}
Using \eqref{inverttheta} we find the field redefinition \eqref{totredef} in terms of $\theta^1$ and $\theta^2$
\begin{equation}
\begin{array}{c} \ds
\Delta \theta^1 =  \Big( - (m + n) + \frac{3}{2} ( m+n)^2 \Big) \theta^1
\\[4mm] \ds
\Delta \theta^2  =  \Big( - \Gamma_{049}(m-n)\Gamma_{049} + \frac{3}{2} ( \Gamma_{049}(m-n)\Gamma_{049} )^2 \Big) \theta^2
\end{array}
\end{equation}
Finally, from \eqref{fermmomenta2} we see that we can write the field redefinition for $\theta$ on the form
\begin{equation}
\label{finalredef}
\begin{array}{c}
\Delta \theta = \Big[ - ( Y + \Gamma_{11} \tilde{Y} ) + \frac{3}{2} ( Y + \Gamma_{11} \tilde{Y} )^2 \Big] \theta
\\[2mm] \ds
Y = \frac{1}{cR} \sum_{i=1}^8 p_i  ( \CP_+ + \frac{1}{2} \CP_- ) \Gamma^0 \Gamma_i ( \CP_+ + \CP_- ) \spa \tilde{Y} =\frac{1}{cR} \sum_{i=1}^8  {X'}^i ( \CP_+ + \frac{1}{2} \CP_- ) \Gamma^0 \Gamma_i ( \CP_+ + \CP_- )
\end{array}
\end{equation}

\subsubsection*{Extra terms induced by the field redefinition}

We now consider what extra terms should be added to the $1/R$ and $1/R^2$ terms of the Hamiltonian \eqref{pH3BF}-\eqref{pH4F} in accordance with the field redefinition \eqref{finalredef}.
Using $\Delta \theta$ with $\CH_2$ we see that we should add the following cubic terms
\begin{equation}
\label{deltaH3BF}
\Delta \CH_{3,BF} = \frac{i}{c} \sum_{i=1}^8 ( - {X'}^i
A_{i,\sigma} + p_i \tilde{A}_{i,\sigma}) +
\frac{i}{4} \sum_{i=1}^8 ( p_i B_{i56} - {X'}^i \tilde{B}_{i56} )
- \frac{i}{2}
\sum_{i=1}^8 ( p_i C_{+i} + {X'}^i \tilde{C}_{+i} )
\end{equation}
For the quartic terms we have three different sources of contributions. Either from $\Delta \theta$ in $\CH_2$, from $\Delta \theta$ twice in $\CH_2$ or from $\Delta \theta$ in $\CH_3$. This gives the following quartic terms that all should be added to $\CH_{4,BF}$
\begin{equation}
\label{deltaH4BF}
\begin{array}{l} \ds
\Delta \CH_{4,BF} = \frac{i}{2 c^2  } \sum_{i,j=1}^8 (p_i p_j' + {X'}^i {X''}^j ) \tilde{E}_{ij} - \frac{i}{2 c^2  } \sum_{i,j=1}^8 ({X'}^i p_j' + p_i {X''}^j ) E_{ij} \\[2mm] \ds - \frac{3i}{4 c  } \sum_{i,j=1}^8 (p_i p_j - {X'}^i {X'}^j ) \CC_{i+;j} + \frac{3i}{4 c  } \sum_{i,j=1}^8 ({X'}^i p_j - p_i {X'}^j ) \tilde{\CC}_{i+;j} \\[2mm] \ds - \frac{i}{4 c  } \sum_{i,j=1}^8 (p_i p_j + {X'}^i {X'}^j ) \CC_{+i;j} - \frac{i}{4 c  } \sum_{i,j=1}^8 ({X'}^i p_j + p_i {X'}^j ) \tilde{\CC}_{+i;j}
\\[2mm] \ds
+
\frac{i}{4c  } \sum_{i,j=5}^8 \sum_{k=1}^8 \epsilon_{ij} \Big[ (p_ip_k + {X'}^i {X'}^k)(\CB_{+-i;k} + E_{jk}) + ({X'}^i
p_k + p_i {X'}^k) (\tilde{\CB}_{+-i;k}-\tilde{E}_{jk}) \Big]
\\[2mm] \ds
+ \frac{i u_4}{2   } \sum_{i=1}^8 ( p_j \CB_{+-4;i} - {X'}^j \tilde{\CB}_{+-4;i} )
+ \frac{iu_4}{2} \sum_{i=1}^8 \Big[ p_i (B_{i56}-B_{i78} ) - {X'}^i (\tilde{B}_{i56} - \tilde{B}_{i78} ) \Big]
\\[2mm] \ds
+  \frac{i}{2c  } \sum_{i=5}^8 \sum_{j=1}^8 s_i \Big[ (p_i p_j - {X'}^i {X'}^j) \CB_{+4i;j} + ({X'}^i p_j - p_i {X'}^j) \tilde{\CB}_{+4i;j}    \Big]
\end{array}
\end{equation}
We defined here the purely fermionic terms
\begin{equation}
\label{calABC}
\begin{array}{c} \ds
\CB_{abc;d} = \bar{\theta} \Gamma_{abc} ( \CP_+ + \frac{1}{2} \CP_- ) \Gamma^0 \Gamma_d \theta \spa \tilde{\CB}_{abc;d} = \bar{\theta} \Gamma_{11} \Gamma_{abc} ( \CP_+ + \frac{1}{2} \CP_- ) \Gamma^0 \Gamma_d \theta
\\[2mm] \ds
\CC_{ab;c} = \bar{\theta} \Gamma_a P \Gamma_{0123} \Gamma_b ( \CP_+ + \frac{1}{2} \CP_- ) \Gamma^0 \Gamma_c \theta \spa \tilde{\CC}_{ab;c} = \bar{\theta} \Gamma_{11} \Gamma_a P \Gamma_{0123} \Gamma_b ( \CP_+ + \frac{1}{2} \CP_- ) \Gamma^0 \Gamma_c \theta
\\[2mm] \ds
E_{ab} = \bar{\theta} \Gamma_a ( \CP_+ + \frac{1}{2} \CP_- ) \Gamma^0 \Gamma_b \theta \spa \tilde{E}_{ab} = \bar{\theta} \Gamma_{11} \Gamma_a ( \CP_+ + \frac{1}{2} \CP_- ) \Gamma^0 \Gamma_b \theta
\end{array}
\end{equation}
%

\subsection{Final expression for fermionic terms in Hamiltonian}

We are now ready to write the final expressions for the fermionic terms in the Hamiltonian. The pp-wave Hamiltonian $\CH_{2,F}$ is given by \eqref{CH2F}, as computed in Section \ref{sec:ppwave}. Combining \eqref{pH3BF} and \eqref{deltaH3BF} we see that the cubic piece $\CH_{3,BF}$ is given by
\begin{eqnarray}
\label{finalH3BF}
\begin{array}{l} \ds
\CH_{3,BF} =  \frac{i}{2} \sum_{i=1}^8 (
 C_{+i}  p_i + \tilde{C}_{+i} {X'}^i )
- \frac{ic}{4} ( B_{+56} - B_{+78} ) u_4
 - \frac{ic}{4}  B_{+-4} u_i
\\[2mm] \ds
  - \frac{i }{4}
\sum_{i=5}^8 s_i ( B_{+4i} p_i + \tilde{B}_{+4i} {X'}^i )
- \frac{i}{8} \sum_{i,j=5}^8 \epsilon_{ij} ( B_{+-i} p_j +
\tilde{B}_{+-i} {X'}^j )
\end{array}
\end{eqnarray}
Combining \eqref{pH4BF} and \eqref{deltaH4BF} we see that the two-boson-two-fermion piece $\CH_{4,BF}$ is given by
\begin{equation}
\label{finalH4BF}
\begin{array}{l} \ds
\CH_{4,BF} = \frac{i}{c^2}
\sum_{i=1}^8 \Big( p_i^2 + ({X'}^i)^2 \Big) \Big[ \tilde{A}_{+,\sigma}
 + \frac{c}{4} ( B_{+56} + B_{+78}  -  C_{++}  +  C_{+-} ) \Big]
- i \tilde{A}_{+,\sigma} \Big[ \sum_{i=1}^3 u_i^2 -u_4^2 \Big]
\\[2mm] \ds
+ \frac{2i}{c^2} \sum_{i=1}^8 p_i  {X'}^i \Big[  A_{+,\sigma} +
\frac{c}{4} ( \tilde{B}_{+56}+\tilde{B}_{+78}) + \frac{c}{4}
\tilde{C}_{+-}   \Big]
+ \frac{ic}{2} \sum_{i=1}^3 u_i^2 C_{++}
- \frac{ic}{4} \sum_{i=1}^4 u_i^2 ( B_{+56} + B_{+78})
\\[2mm] \ds
+ \frac{i}{2} u_4 \sum_{i=5}^8 s_i \Big[ C_{+i}  p_i  -
\tilde{C}_{+i} {X'}^i \Big]
- \frac{i}{c} \sum_{i,j=1}^8 \Big[  C_{ij} ({X'}^i {X'}^j - p_i  p_j
) + 2  \tilde{C}_{ij} {X'}^i p_j  \Big]
- i \sum_{i,j=1}^3 u_i' u_j \tilde{B}_{+ij}
\\[2mm] \ds
- \frac{i}{8} u_4 \sum_{i,j=5}^8 s_i \epsilon_{ij} ( 3 B_{+-i} p_j
+ \tilde{B}_{+-i} {X'}^j )
+ \frac{i}{4} ( B_{+56} p_{x_1} y_1 +  \tilde{B}_{+56} x_1' y_1 +  B_{+78} p_{x_2} y_2 +  \tilde{B}_{+78} x_2' y_2 )
\\[2mm] \ds
- \frac{i}{2} \sum_{i=1}^4 \sum_{j=1}^8 u_i \Big[ B_{-ij} p_j  - \tilde{B}_{-ij} {X'}^j \Big]
+ \frac{i}{2c} \sum_{i=1}^8 \sum_{j=5}^8 s_j \Big[ ( p_i  p_j  - {X'}^i {X'}^j ) B_{4 i j} + ( p_i  {X'}^j - {X'}^i p_j  ) \tilde{B}_{4 i j} \Big]
\\[2mm] \ds
- \frac{i}{2} \sum_{i=1}^3 \sum_{j=4}^8 u_i \Big[ B_{+ij} p_j -  \tilde{B}_{+ij} {X'}^j \Big]
- \frac{i}{4} u_4 \sum_{i=5}^8 ( B_{+4i} p_i  + 3 \tilde{B}_{+4i} {X'}^i )
+ \frac{i}{2} u_4 \sum_{i=1}^3 ( B_{+4i} p_i - \tilde{B}_{+4i} u_i' )
\\[2mm] \ds
- \frac{i}{4c} \sum_{i=1}^8 \sum_{j,k=5}^8 \epsilon_{jk} \Big[
(B_{+ij} - B_{-ij}) ( p_i  p_k  - {X'}^i {X'}^k )  +
(\tilde{B}_{+ij}-\tilde{B}_{-ij}) ( p_i  {X'}^k - {X'}^i p_k  )
\Big]
\\[2mm] \ds + \frac{i}{2 c^2  } \sum_{i,j=1}^8 (p_i p_j' + {X'}^i {X''}^j ) \tilde{E}_{ij} - \frac{i}{2 c^2  } \sum_{i,j=1}^8 ({X'}^i p_j' + p_i {X''}^j ) E_{ij} - \frac{3i}{4 c  } \sum_{i,j=1}^8 (p_i p_j - {X'}^i {X'}^j ) \CC_{i+;j}  %
\\[2mm] \ds
+ \frac{3i}{4 c  } \sum_{i,j=1}^8 ({X'}^i p_j - p_i {X'}^j ) \tilde{\CC}_{i+;j} - \frac{i}{4 c  } \sum_{i,j=1}^8 (p_i p_j + {X'}^i {X'}^j ) \CC_{+i;j} - \frac{i}{4 c  } \sum_{i,j=1}^8 ({X'}^i p_j + p_i {X'}^j ) \tilde{\CC}_{+i;j}
\\[2mm] \ds
+ \frac{i u_4}{2   } \sum_{i=1}^8 ( p_j \CB_{+-4;i} - {X'}^j \tilde{\CB}_{+-4;i} )
+  \frac{i}{2c  } \sum_{i=5}^8 \sum_{j=1}^8 s_i \Big[ (p_i p_j - {X'}^i {X'}^j) \CB_{+4i;j} + ({X'}^i p_j - p_i {X'}^j) \tilde{\CB}_{+4i;j}    \Big]
\\[2mm] \ds
+
\frac{i}{4c  } \sum_{i,j=5}^8 \sum_{k=1}^8 \epsilon_{ij} \Big[ (p_ip_k + {X'}^i {X'}^k)(\CB_{+-i;k} + E_{jk}) + ({X'}^i
p_k + p_i {X'}^k) (\tilde{\CB}_{+-i;k}-\tilde{E}_{jk}) \Big]
\end{array}
\end{equation}
Here we still use the definitions \eqref{ABCdef} and \eqref{calABC} as the definitions of the various two-fermion objects. However, we take $\theta$ in these definitions to be given in terms of $\psi$ and $\rho$ as
\begin{equation}
\label{thetheta}
\theta (\psi,\rho) = \frac{1}{2} ( \psi + E^{-1} \rho ) + \frac{\Gamma_{049}}{2i} ( \psi - E^{-1} \rho )
\end{equation}
In this sense we have specified how $\CH_{3,BF}$ and $\CH_{4,BF}$ in \eqref{finalH3BF}-\eqref{finalH4BF} depends on the fermionic phase space variables $\psi$ and $\rho$.
Finally, the four-fermion Hamiltonian $\CH_{4,F}$ is given by \eqref{pH4F} since the field redefinition \eqref{finalredef} does not induce any additional four-fermion terms. Again, the two-fermion objects in \eqref{pH4F} are defined by \eqref{ABCdef} with $\theta (\psi,\rho)$ given by \eqref{thetheta}.

Thus, we have now specified all the quadratic, cubic and quartic terms in the Hamiltonian in terms of the bosonic and fermionic phase space variables. And, furthermore, by implementing the Dirac procedure as part of a field redefinition we have made sure that the quantized Hamiltonian has the canonical commutation relations for the fermions, $i.e.$ that the anti-commutation relation \eqref{anticommu} is not corrected by $1/R$ or $1/R^2$ corrections.

\section*{Acknowledgments}

We thank M. Cvetic, P. A. Grassi, S. Hirano, C. Pope, R. Roiban, K. Stelle, A. Tseytlin,  K. Zarembo, K.  Zoubos, and especially D. Sorokin for many interesting and stimulating discussions and correspondence. We thank the Referee of Journal of High-Energy Physics for useful comments and suggestions to the manuscript. GG and MO thank the Galileo Galilei Institute for Theoretical Physics for hospitality and the INFN for partial support during the completion of this work. The work of GG is partially supported by the MIUR-PRIN contract 2007-5ATT78.

\begin{appendix}

\section{$\ads_4 \times \C P^3$ background}
\label{app:background}

The $\ads_4\times \C P^3$ background has the metric
\begin{equation}
\label{adscp} ds^2 = \frac{R^2}{4} \left( - \cosh^2 \rho dt^2 +
d\rho^2 + \sinh^2 \rho d\hat{\Omega}^2_2 \right) + R^2 ds_{\C P^3}^2
\end{equation}
with
\begin{equation}
\label{cp32} ds_{\C P^3}^2 = \frac{1}{4} d\psi^2 + \frac{1-\sin
\psi}{8} d\Omega_2^2 + \frac{1+\sin \psi}{8} d{\Omega_2'}^2 + \cos^2
\psi ( d \delta + \omega )^2
\end{equation}
where the one-form $\omega$ and the metrics for the two two-spheres
are given by
\begin{equation}
\omega = \frac{1}{4} \sin \theta_1 d\varphi_1 + \frac{1}{4} \sin
\theta_2 d\varphi_2
\end{equation}
\begin{equation}
d\Omega_2^2 = d\theta_1^2 + \cos^2 \theta_1 d\varphi_1^2 \spa
{d\Omega_2'}^2 = d\theta_2^2 + \cos^2 \theta_2 d\varphi_2^2
\end{equation}

The two-form and four-form field strengths are
\begin{equation}
\frac{1}{R} F_{(2)} = - \cos \psi d\psi \wedge ( d\delta + \omega) +
\frac{1-\sin \psi}{4} \cos \theta_1 d\theta_1 \wedge d\varphi_1 -
\frac{1+\sin \psi}{4} \cos \theta_2 d\theta_2 \wedge d\varphi_2
\end{equation}
\begin{equation}
\frac{1}{R^3} F_{(4)} = \frac{3}{8} \epsilon_{\ads_4} = \frac{3}{8}
\cosh \rho \sinh^2 \rho dt \wedge d\rho \wedge d\hat{\Omega}_2
\end{equation}

The curvature radius $R$ is given by
\begin{equation}
R^4 = 32 \pi^2 \lambda l_s^4
\end{equation}

Defining the coordinates $u_1$, $u_2$ and $u_3$ by
\begin{equation}
\frac{R}{2} \sinh \rho = \frac{u}{1 - \frac{u^2}{R^2}} \spa
\frac{R^2}{4} ( d\rho^2 + \sinh^2 \rho d\hat{\Omega}_2^2 ) =
\frac{\sum_{i=1}^3 du_i^2}{(1-\frac{u^2}{R^2}  )^2} \spa u^2 =
\sum_{i=1}^3 u_i^2
\end{equation}
and the coordinates $x_i,y_i$, $i=1,2$, and $u_4$ by
\begin{equation}
x_1 = R \varphi_1 \spa y_1 = R \theta_1 \spa x_2 =
R \varphi_2 \spa y_2 = R \theta_2 \spa u_4 = \frac{R}{2} \psi
\end{equation}
we can write the metric as
\begin{eqnarray}
ds^2  & = & - \frac{R^2}{4} \frac{(1+\frac{u^2}{R^2}  )^2}{(1-\frac{u^2}{R^2}  )^2} dt^2 +
\frac{\sum_{i=1}^3 du_i^2}{(1-\frac{u^2}{R^2}  )^2}
+ du_4^2 +R^2\cos^2\frac{2u_4}{R} \left[ d\delta^2 + 2 d\delta \left( \sin
\frac{y_1}{R} \frac{d x_1}{4R} + \sin \frac{y_2}{R} \frac{d x_2}{4R} \right)
\right] \nn \\
& &
+\frac{1}{8}\left(\cos\frac{u_4}{R}-\sin\frac{u_4}{R}\right)^2\left(
dy_1^2+\cos^2\frac{y_1}{R}d
x_1^2\right)
+\frac{1}{8}\left(\cos\frac{u_4}{R}+\sin\frac{u_4}{R}\right)^2\left(
dy_2^2+\cos^2\frac{y_2}{R}d x_2^2\right)
\nn \\ & &
\end{eqnarray}
This corresponds to the zehnbeins
\begin{equation}
\label{zehnbeins1} e^{0} = \frac{R}{2}
\frac{1+\frac{u^2}{R^2}}{1-\frac{u^2}{R^2}} dt \spa e^i =
\frac{du_i}{1-\frac{u^2}{R^2}}  \ , \ i=1,2,3
\end{equation}
\begin{equation}
\label{zehnbeins2} e^5 = \frac{1}{2\sqrt{2}} \left( \cos
\frac{u_4}{R} - \sin \frac{u_4}{R} \right) \cos \frac{y_1}{R} dx_1
\spa e^6 = \frac{1}{2\sqrt{2}} \left( \cos \frac{u_4}{R} - \sin
\frac{u_4}{R} \right) dy_1
\end{equation}
\begin{equation}
\label{zehnbeins3} e^7 = \frac{1}{2\sqrt{2}} \left( \cos
\frac{u_4}{R} + \sin \frac{u_4}{R} \right) \cos \frac{y_2}{R} dx_2
\spa e^8 = \frac{1}{2\sqrt{2}} \left( \cos \frac{u_4}{R} + \sin
\frac{u_4}{R} \right) dy_2
\end{equation}
\begin{equation}
\label{zehnbeins4} e^4 = du_4 \spa e^9 = \frac{R}{2} \cos
\frac{2u_4}{R} \left[ 2 d \delta + \frac{1}{2R } \left( \sin
\frac{y_1}{R} dx_1 + \sin \frac{y_2}{R} dx_2 \right) \right]
\end{equation}
Using these the two-form and four-form field strengths takes the form
\begin{equation}
\label{F2bein} F_{(2)} = \frac{2}{R} ( - e^4 \wedge e^9 - e^5 \wedge
e^6 + e^7 \wedge e^8 )
\end{equation}
\begin{equation}
\label{F4bein} F_{(4)} = \frac{6}{R} e^0 \wedge e^1 \wedge e^2
\wedge e^3
\end{equation}

We make the coordinate transformation
\begin{equation}
\delta = \frac{1}{2} t + \frac{v}{R^2}
\end{equation}
Written explicitly, the metric in these coordinates
becomes
\begin{align}
&ds^2 = -  dt^2 \left( \frac{R^2}{4} \sin^2
\frac{2u_4}{R} + \frac{u^2}{(1 - \frac{u^2}{R^2})^2} \right) +
\frac{\sum_{i=1}^3 du_i^2}{(1-\frac{u^2}{R^2}  )^2} + du_4^2\nn \\
&
+\frac{1}{8}\left(\cos\frac{u_4}{R}-\sin\frac{u_4}{R}\right)^2\left(
dy_1^2+\cos^2\frac{y_1}{R}d
x_1^2\right)+\frac{1}{8}\left( \cos\frac{u_4}{R}+\sin\frac{u_4}{R}\right)^2\left(
dy_2^2+\cos^2\frac{y_2}{R}d x_2^2\right)
\cr&+R^2\cos^2\frac{2u_4}{R} \left[ dt + \frac{dv}{R^2} + \sin
\frac{y_1}{R} \frac{d x_1}{4R} + \sin \frac{y_2}{R} \frac{d x_2}{4R}
\right] \left[ \frac{dv}{R^2} + \sin \frac{y_1}{R} \frac{d x_1}{4R}
+ \sin \frac{y_2}{R} \frac{d x_2}{4R} \right]
\end{align}

Define
\begin{equation}
e^+ = \frac{1}{2R} ( e^0 + e^9 ) \spa e^- = \frac{R}{2} ( e^0 - e^9
)
\end{equation}

\section{Gamma-matrix conventions}
\label{app:gamma}

Define the real $8 \times 8$ matrices $\gamma_1,...,\gamma_8$ as in
\cite{Callan:2004uv}. They obey
\begin{equation}
\begin{array}{c} \ds
\gamma_i \gamma_j^T + \gamma_j \gamma_i^T = \gamma_i^T \gamma_j +
\gamma_j^T \gamma_i = 2 \delta_{ij} I_8 \, , \, i,j=1,...,8
\\[4mm] \ds
\gamma_1 \gamma_2^T \gamma_3 \gamma_4^T \gamma_5 \gamma_6^T \gamma_7
\gamma_8^T = I_8 \spa \gamma_1^T \gamma_2 \gamma_3^T \gamma_4
\gamma_5^T \gamma_6 \gamma_7^T \gamma_8 = - I_8
\end{array}
\end{equation}
where $I_n$ is the $n\times n$ identity matrix. Define the $16
\times 16$ matrices $\hat{\gamma}_1,...,\hat{\gamma}_9$ by
\begin{equation}
\hat{\gamma}_i = \matrto{0}{\gamma_i}{\gamma_i^T}{0}\, , \,
i=1,...,8 \spa \hat{\gamma}_{9} = \matrto{I_8}{0}{0}{-I_8}
\end{equation}
The matrices $\hat{\gamma}_1,...,\hat{\gamma}_9$ are symmetric and
real and they obey
\begin{equation}
\{ \hat{\gamma}_i, \hat{\gamma}_j \} = 2 \delta_{ij} I_{16} \, , \,
i,j=1,...,9 \spa \hat{\gamma}_9 = \hat{\gamma}_1 \hat{\gamma}_2
\cdots \hat{\gamma}_8
\end{equation}
Define the $32 \times 32$ matrices
\begin{equation}
\Gamma_0  = \matrto{0}{-I_{16}}{I_{16}}{0} \spa \Gamma_i   =
\matrto{0}{\hat{\gamma}_i}{\hat{\gamma}_i}{0} \, , \, i=1,...,9 \spa
\Gamma_{11}  = \matrto{I_{16}}{0}{0}{-I_{16}}
\end{equation}
These matrices are real and obey
\begin{equation}
\{ \Gamma_a,\Gamma_b \} = 2 \eta_{ab} I_{32} \, , \,
i,j=0,1,...,9,11 \spa \Gamma_{11} = \Gamma^0 \Gamma^1 \cdots
\Gamma^9
\end{equation}
We define
\begin{equation}
\gamma_{i_1 \cdots i_{2k} } = \gamma_{[i_1} \gamma^T_{i_2}\cdots
\gamma^T_{i_{2k}]} \spa \gamma_{i_1 i_2 \cdots i_{2k+1} } =
\gamma^T_{[i_1} \gamma_{i_2} \cdots \gamma^T_{i_{2k+1}]} \spa i_l =
1,...,8
\end{equation}
\begin{equation}
\hat{\gamma}_{i_1 \cdots i_n } = \hat{\gamma}_{[i_1}
\hat{\gamma}_{i_2} \cdots \hat{\gamma}_{i_n]} \spa i_l = 1,...,9
\end{equation}
\begin{equation}
\Gamma_{i_1 i_2 \cdots i_n} = \Gamma_{[i_1} \Gamma_{i_2} \cdots
\Gamma_{i_n]} \spa i_l = 0,1,...,9,11
\end{equation}
%

\section{Structure constants and $\mathcal{M}^2$}
\label{app:Msqr}

We can write the $OSp(6|2,2)$ algebra schematically on the form
\begin{equation}
\label{fullalg} [B_i,B_j] = f^k_{ij} B_k \spa [F_\alpha,B_i] =
\tilde{f}^\beta_{\alpha i} F_\beta \spa \{ F_\alpha,F_\beta \} =
\hat{f}^i_{\alpha\beta} B_i
\end{equation}
where $B_i$ are the bosonic generators which generate an $SO(2,3)
\times SU(4)$ algebra (with $25=10+15$ generators) and $F_\alpha$ corresponds to the 24 fermionic generators. We take $\alpha$ to run over all $32$ fermionic directions. The 24 fermionic generators are then defined by $P_{\alpha\beta} F_\beta= F_\alpha$.

\subsubsection*{The structure constants}

The structure constants $\tilde{f}^\beta_{\alpha i}$ can be read off from the covariant derivative \eqref{covdertheta}%
\footnote{Note that $\tilde{f}^\alpha_{\beta a}$ for $a=0,1,2,3$ corresponds to four of the $SO(2,3)$ generators, the other six corresponding to $\tilde{f}^\alpha_{\beta\,\hat{a}\hat{b}}$ since this is antisymmetric in $\hat{a},\hat{b}$. Similarly, the $\tilde{f}^\alpha_{\beta a}$ for $a=4,5,...,9$ corresponds to six of the $SU(4)$ generators. This leaves the last 9 for $\tilde{f}^\alpha_{\beta\,a'b'}$, despite the fact that antisymmetry of $a',b'$ seemingly gives 15. However, the projector $P$ in $\tilde{f}^\alpha_{\beta\,a'b'}$ gives relations between the matrices therefore only 9 of them are independent. Thus, we get the 15 generators of $SU(4)$. The same story is true for $\hat{f}^i_{\alpha\beta}$.}
\begin{equation}
\label{theftildes}
\tilde{f}^\alpha_{\beta a} = \frac{1}{R} ( \Gamma_{0123} P \Gamma_a P )^\alpha_{\ \beta}
\spa
\tilde{f}^\alpha_{\beta\,\hat{a}\hat{b}} = - \frac{1}{4} ( P \Gamma_{\hat{a}\hat{b}} P )^\alpha_{\ \beta}
\spa
\tilde{f}^\alpha_{\beta\,a'b'} = - \frac{1}{4} ( P \Gamma_{a'b'} P )^\alpha_{\ \beta}
\end{equation}
where $a=0,1,...,9$, $\hat{a},\hat{b}=0,1,2,3$ and $a',b'=4,5,...,9$.
The structure constants $\hat{f}^i_{\alpha\beta}$ are instead
\begin{equation}
\label{thefhats}
\begin{array}{c} \ds
\hat{f}^a_{\alpha\beta} =2 i ( P \Gamma^0
\Gamma^a P )_{\alpha\beta}
\\[3mm] \ds \hat{f}^{\hat{a}\hat{b}}_{\alpha\beta}
  =-\frac{4i}{R}(P\Gamma^0\Gamma_{0123}\Gamma^{\hat{a}\hat{b}}P)_{\alpha\beta} \\[3mm] \ds
  \hat{f}^{{a'}{b'}}_{\alpha\beta} = \frac{2i}{R}\left(P\Gamma^0\left(\Gamma_{0123}\Gamma^{a'b'}-J^{a'b'}\Gamma_{11}\right)P\right)_{\alpha\beta}
\end{array}
\end{equation}
where $a=0,1,...,9$, $\hat{a},\hat{b}=0,1,2,3$ and $a',b'=4,5,...,9$, and where we introduced the Kaehler form
\begin{equation}\label{theJab}
   J^{a'b'}=\delta^{a'7}\delta^{b'8}-\delta^{a'8}\delta^{b'7}-\delta^{a'4}\delta^{b'9}
   +\delta^{a'9}\delta^{b'4}-\delta^{a'5}\delta^{b'6}+\delta^{a'6}\delta^{b'5}
\end{equation}
For $a'=4,\dots,9$ the structure constants $\hat{f}^{a'}_{\alpha\beta}$ can also be written as
\begin{equation}\label{thefhat} \hat{f}^{a'}_{\alpha\beta} = i(P \Gamma^0\Gamma^{a'}P)_{\alpha\beta}+i(P
\Gamma^0\Gamma_{0123}\Gamma_{11}J^{a' b'}\Gamma_{b'}P)_{\alpha\beta}
\end{equation}
using here the relation
\begin{equation}\label{theprop}
(P\Gamma^{a'}P)_{\alpha\beta}=(P\Gamma_{0123}\Gamma_{11}J^{a'b'}\Gamma_{b'}P)_{\alpha\beta}
\end{equation}

\subsubsection*{The fermionic matrix $\CM^2$}

We now determine the fermionic matrix $\CM^2$ needed for writing the four-fermion terms in the Lagrangian of Section \ref{sec:genlagr}. It can generally be written in terms of structure constants of the $OSp(6|2,2)$ algebra \eqref{fullalg} as
\begin{equation}
(\mathcal{M}^2)^\alpha_\beta = - \theta^\gamma \tilde{f}^\alpha_{\gamma i} \theta^\delta \hat{f}^i_{\delta \beta}=-
\theta^\gamma \tilde{f}^\alpha_{\gamma a} \theta^\delta \hat{f}^a_{\delta \beta}- \theta^\gamma
\tilde{f}^\alpha_{\gamma\, ab} \theta^\delta \hat{f}^{ab}_{\delta \beta}
\end{equation}
Using the structure constants written above we compute
\begin{eqnarray}
&&\left(\mathcal{M}^2\right)_\beta^\alpha =-\frac{2i}{R}(P\Gamma_{0123}\Gamma_a P)^{\alpha}_{\
\gamma}\theta^{\gamma}\theta^{\delta}(P\Gamma^{0}\Gamma^a P)_{\delta \beta}
\cr&&-\frac{i}{R}(P\Gamma_{\hat{a}\hat{b}} P)^{\alpha}_{\
\gamma}\theta^{\gamma}\theta^{\delta}(P\Gamma^{0}\Gamma_{0123}\Gamma^{\hat{a}\hat{b}} P)_{\delta
\beta}\cr&&+\frac{i}{2R}(P\Gamma_{{a'}{b'}} P)^{\alpha}_{\
\gamma}\theta^{\gamma}\theta^{\delta}(P\Gamma^{0}(\Gamma_{0123}\Gamma^{{a'}{b'}}-J^{{a'}{b'}}\Gamma_{11})
P)_{\delta \beta}
\end{eqnarray}
where $\hat{a},\hat{b}=0,1,2,3$ and $a',b'=4,5,...,9$. This can also be written in the form
\begin{eqnarray}
\label{them2formula}
\left(\mathcal{M}^2\right)_\beta^\alpha &=&-\frac{2i}{R}(P\Gamma_{0123}\Gamma_{\hat{a}}
P)^{\alpha}_{\ \gamma}\theta^{\gamma}\theta^{\delta}(P\Gamma^{0}\Gamma^{\hat{a}} P)_{\delta
\beta} - \frac{i}{R}(P\Gamma_{0123}\Gamma_{{a'}} P)^{\alpha}_{\
\gamma}\theta^{\gamma}\theta^{\delta}(P\Gamma^{0}\Gamma^{{a'}} P)_{\delta
\beta}\cr&+&\frac{i}{R}(P\Gamma_{11}\Gamma_{{a'}} P)^{\alpha}_{\
\gamma}\theta^{\gamma}\theta^{\delta}(P\Gamma^{0}\Gamma_{0123}\Gamma_{11}\Gamma^{{a'}} P)_{\delta
\beta} - \frac{i}{R}(P\Gamma_{\hat{a}\hat{b}} P)^{\alpha}_{\
\gamma}\theta^{\gamma}\theta^{\delta}(P\Gamma^{0}\Gamma_{0123}\Gamma^{\hat{a}\hat{b}} P)_{\delta
\beta}\cr &+&\frac{i}{2R}(P\Gamma_{{a'}{b'}} P)^{\alpha}_{\
\gamma}\theta^{\gamma}\theta^{\delta}(P\Gamma^{0}\Gamma_{0123}\Gamma^{{a'}{b'}} P)_{\delta
\beta} - \frac{i}{R}(\Gamma_{0123}\Gamma_{11} )^{\alpha}_{\
\gamma}\theta^{\gamma}\theta^{\delta}(\Gamma^{0}\Gamma_{11})_{\delta \beta}\cr&&
\end{eqnarray}
where we eliminated $J^{a'b'}$ by taking into account that
\begin{equation}\label{theJ}
    J=\frac{1}{2}\Gamma_{0123}\Gamma_{11}J^{a'b'}\Gamma_{{a'}{b'}}
\end{equation}
along with the relation \eqref{theprop} and that $J$ on supersymmetric fermions gives $J\theta=-\theta$.

\subsubsection*{Equivalence with $\CM^2$ in alternative representation}

We now show that the formula \eqref{them2formula} is equivalent to the one written in
\cite{Gomis:2008jt,Grassi:2009yj}. Thus, we shall use the following alternative representation of the Gamma
matrices \cite{Gomis:2008jt,Grassi:2009yj}
\begin{eqnarray}\label{theGamma10}
&&\Gamma^{\hat a}=\gamma^{\hat a}\,\otimes\,{\bf 1},\qquad
\Gamma^{a'}=\gamma^5\,\otimes\,\gamma^{a'},\qquad \Gamma^{11}=\gamma^5\,\otimes\,\gamma^7,\qquad\cr
&& \hat{a}=0,1,2,3;\quad a'=4,\cdots,9\,. \nonumber
 \end{eqnarray}
Here $(\gamma^{\hat{a}})_{\hat{\alpha}\hat{\beta}}$ are 4-dimensional matrices, corresponding to the $AdS_4$ part,
$\hat{\alpha},\hat{\beta}=1,\dots,4$ and $(\gamma^{a'})_{{\alpha'}{\beta '}}$ are 8-dimensional
matrices,  ${\alpha'}, {\beta '}=1,\dots,8$, corresponding to the 6-dimensional space $\C P^3$.
Eq.\eqref{theprop} becomes
\begin{equation}\label{therelation}
P_6\gamma^{a'}P_6=i P_6\,J^{a'b'}\gamma_{b'}\gamma^7\, P_6
\end{equation}
this was derived in~\cite{Gomis:2008jt}. Here $P_6$ is the reduction of $P$ to $\C P^3$
\begin{equation}\label{P6}
P_6=\frac{3-J}{4}~,~~~~~2 J=-i J^{a'b'}\gamma_{a'b'}\gamma^7~.
\end{equation}
This projector when acting on an 8--dimensional spinor annihilates 2 and leaves 6 of its components. Thus the spinor
\begin{equation}\label{24} \theta^{\hat{\alpha}\alpha'}=({P}_6\,\theta)^{\hat{\alpha}\alpha'} \qquad
\Longleftrightarrow \qquad \theta^{\hat{\alpha} a'}\, \qquad a'=1,\cdots, 6 \end{equation} has 24
non--zero components.
In terms of the dimensionally reduced $\gamma$-matrices $\mathcal{M}^2$
reads
\begin{eqnarray}
 && \left(\mathcal{M}^2\right)_\beta^\alpha =-\frac{2}{R}\left(\gamma^5\gamma_{\hat{a}}\right)^{\hat{\alpha}}_{\ \hat{\gamma}}
  \theta^{\hat{\gamma}a'}\theta^{\hat{\delta}}_{\ b'}(\gamma^0\gamma^{\hat{a}})_{\hat{\delta}\hat{\beta}}
  -\frac{1}{R}(\gamma_{c'})^{a'}_{\ d'}
  \theta^{\hat{\alpha}d'}\theta^{\hat{\delta}}_{\ f'}(\gamma^0\gamma^5)_{\hat{\delta}\hat{\beta}}(\gamma^{c'})^{f'}_{\ b'}  \cr
 &-&\frac{i}{R}(\gamma_{c'})^{a'}_{\ d'}
  \theta^{\hat{\alpha}d'}\theta^{\hat{\delta}}_{\ f'}(\gamma^0\gamma^5)_{\hat{\delta}\hat{\beta}}(J^{c' g'}\gamma_{g'}\gamma^7)^{f'}_{\ b'} -\frac{1}{R}(\gamma_{\hat{a}\hat{b}})^{\hat{\alpha}}_{\ \hat{\gamma}}
  \theta^{\hat{\gamma}a'}\theta^{\hat{\delta}}_{\ b'}(\gamma^0\gamma^5\gamma^{\hat{a}\hat{b}})_{\hat{\delta}\hat{\beta}}\cr
 &+&\frac{1}{2 R}(\gamma_{c' g'})^{a'}_{\ d'}
  \theta^{\hat{\alpha}d'}\theta^{\hat{\delta}}_{\ f'}(\gamma^0\gamma^5)_{\hat{\delta}\hat{\beta}}(\gamma^{c' g'})^{f'}_{\ b'}
  -\frac{i}{2 R}(\gamma_{c' g'})^{a'}_{\ d'}
  \theta^{\hat{\alpha}d'}\theta^{\hat{\delta}}_{\ f'}(\gamma^0\gamma^5)_{\hat{\delta}\hat{\beta}}(J^{c' g'}\gamma^7)^{f'}_{\ b'}\cr&&
\end{eqnarray}
Using the relations
\begin{equation}
\gamma_{c'} J^{c' b'}=i \gamma^{g'}\gamma^7 \spa J^{c' g'}\gamma_{c' g'}\theta=-2 i \gamma^7\theta
\end{equation}
we find
\begin{eqnarray}\label{M2sor}
\left(\mathcal{M}^2\right)_\beta^\alpha =&-&\frac{2}{R}\left(\gamma^5\gamma_{\hat{a}}\right)^{\hat{\alpha}}_{\ \hat{\gamma}}
  \theta^{\hat{\gamma}a'}\theta^{\hat{\delta}}_{\ b'}(\gamma^0\gamma^{\hat{a}})_{\hat{\delta}\hat{\beta}}
  -\frac{1}{R}(\gamma_{\hat{a}\hat{b}})^{\hat{\alpha}}_{\ \hat{\gamma}}
  \theta^{\hat{\gamma}a'}\theta^{\hat{\delta}}_{\ b'}(\gamma^0\gamma^5\gamma^{\hat{a}\hat{b}})_{\hat{\delta}\hat{\beta}}\cr
 &-&\frac{1}{R}\left[(\gamma_{c'})^{a'}_{\ d'} (\gamma^{c'})^{f'}_{\ b'}-(\gamma^7\gamma^{g'})^{a'}_{\ d'}
 (\gamma^7\gamma^{g'})^{f'}_{\ b'} -\frac{1}{2}(\gamma_{c' g'})^{a'}_{\ d'}(\gamma^{c' g'})^{f'}_{\ b'}\right.\cr
 &+&\left. (\gamma^7)^{a'}_{\ d'}(\gamma^7)^{f'}_{\ b'}\right]\theta^{\hat{\alpha}d'}
 \theta^{\hat{\delta}}_{\ f'}(\gamma^0\gamma^5)_{\hat{\delta}\hat{\beta}}
\end{eqnarray}
We can now use the Fierz identity for the 8 dimensional gamma matrices $\gamma^{a'}$ in 6 dimensions
\begin{eqnarray}\label{theFierz}
(\gamma_{a'})_{\alpha\beta}(\gamma^{a'})_{\gamma \delta}&=&4\left(\delta_{\alpha \delta}\delta_{\beta \gamma}-\delta_{\alpha \gamma}\delta_{\beta \delta}\right)+\frac{1}{2}(\gamma_{a' b'})_{\alpha \beta}
(\gamma^{a' b'})_{\gamma \delta}\cr
 &+& (\gamma^7\gamma_{a'})_{\alpha \beta}(\gamma^7\gamma^{a'})_{\gamma \delta}
 - (\gamma^7)_{\alpha \beta} (\gamma^7)_{\gamma \delta}
\end{eqnarray}
giving
\begin{eqnarray}
\left(\mathcal{M}^2\right)_\beta^\alpha
=&-&\frac{2}{R}\left(\gamma^5\gamma_{\hat{a}}\right)^{\hat{\alpha}}_{\ \hat{\gamma}}
  \theta^{\hat{\gamma}a'}\theta^{\hat{\delta}}_{\ b'}(\gamma^0\gamma^{\hat{a}})_{\hat{\delta}\hat{\beta}}
  -\frac{1}{R}(\gamma_{\hat{a}\hat{b}})^{\hat{\alpha}}_{\ \hat{\gamma}}
  \theta^{\hat{\gamma}a'}\theta^{\hat{\delta}}_{\ b'}(\gamma^0\gamma^5\gamma^{\hat{a}\hat{b}})_{\hat{\delta}\hat{\beta}}\cr
&+&\frac{4}{R}\theta^{\hat{a}}_{\ b'}(\theta^{a'}\gamma^5)_{\hat{\beta}}- \frac{4}{R}\delta^{a'}_{\
b'}\theta^{\hat{a} c'}(\theta \gamma^5)_{\hat{\beta} c'}
\end{eqnarray}
which is the $\mathcal{M}^2$ found in Ref.\cite{Gomis:2008jt,Grassi:2009yj}.

\end{appendix}

\small


\begin{thebibliography}{10}

\bibitem{Maldacena:1997re}
J.~M. Maldacena, ``{The large N limit of superconformal field theories and
  supergravity},'' {\em Adv. Theor. Math. Phys.} {\bf 2} (1998) 231--252,
\href{http://www.arXiv.org/abs/hep-th/9711200}{{\tt hep-th/9711200}}.

\bibitem{Schwarz:2004yj}
J.~H. Schwarz, ``{Superconformal Chern-Simons theories},'' {\em JHEP} {\bf 11}
  (2004) 078,
\href{http://www.arXiv.org/abs/hep-th/0411077}{{\tt hep-th/0411077}}.
%
D.~Gaiotto and X.~Yin, ``{Notes on superconformal Chern-Simons-matter
  theories},'' {\em JHEP} {\bf 08} (2007) 056,
\href{http://www.arXiv.org/abs/0704.3740}{{\tt 0704.3740}}.

\bibitem{Aharony:2008ug}
O.~Aharony, O.~Bergman, D.~L. Jafferis, and J.~Maldacena, ``{$\CN=6$}
  superconformal {Chern-Simons-matter} theories, {M2-branes} and their gravity
  duals,''
\href{http://www.arXiv.org/abs/0806.1218}{{\tt 0806.1218}}.

\bibitem{Minahan:2008hf}
J.~A. Minahan and K.~Zarembo, ``{The Bethe ansatz for superconformal
  Chern-Simons},'' {\em JHEP} {\bf 09} (2008) 040,
\href{http://www.arXiv.org/abs/0806.3951}{{\tt 0806.3951}}.

\bibitem{Gaiotto:2008cg}
D.~Gaiotto, S.~Giombi, and X.~Yin, ``{Spin Chains in N=6 Superconformal
  Chern-Simons-Matter Theory},'' {\em JHEP} {\bf 04} (2009) 066,
\href{http://www.arXiv.org/abs/0806.4589}{{\tt 0806.4589}}.

\bibitem{Arutyunov:2008if}
G.~Arutyunov and S.~Frolov, ``{Superstrings on {$\ads_4 \times \C P^3$} as a
  Coset Sigma-model},'' {\em JHEP} {\bf 09} (2008) 129,
\href{http://www.arXiv.org/abs/0806.4940}{{\tt 0806.4940}}.

\bibitem{Stefanski:2008ik}
B.~Stefanski, jr, ``{Green-Schwarz action for Type IIA strings on $AdS_4\times
  CP^3$},'' {\em Nucl. Phys.} {\bf B808} (2009) 80--87,
\href{http://www.arXiv.org/abs/0806.4948}{{\tt 0806.4948}}.

\bibitem{Grignani:2008is}
G.~Grignani, T.~Harmark, and M.~Orselli, ``{The SU(2) x SU(2) sector in the
  string dual of N=6 superconformal Chern-Simons theory},'' {\em Nucl. Phys.}
  {\bf B810} (2009) 115--134,
\href{http://www.arXiv.org/abs/0806.4959}{{\tt 0806.4959}}.

\bibitem{Gromov:2008qe}
N.~Gromov and P.~Vieira, ``{The all loop AdS4/CFT3 Bethe ansatz},'' {\em JHEP}
  {\bf 01} (2009) 016,
\href{http://www.arXiv.org/abs/0807.0777}{{\tt 0807.0777}}.

\bibitem{Astolfi:2008ji}
D.~Astolfi, V.~G.~M. Puletti, G.~Grignani, T.~Harmark, and M.~Orselli,
  ``{Finite-size corrections in the SU(2) $\times$ SU(2) sector of type IIA
  string theory on {$\ads_4 \times \C P^3$}},'' {\em Nucl. Phys.} {\bf B810}
  (2009) 150--173,
\href{http://www.arXiv.org/abs/0807.1527}{{\tt 0807.1527}}.

\bibitem{Bak:2008cp}
D.~Bak and S.-J. Rey, ``{Integrable Spin Chain in Superconformal Chern-Simons
  Theory},'' {\em JHEP} {\bf 10} (2008) 053,
\href{http://www.arXiv.org/abs/0807.2063}{{\tt 0807.2063}}.

\bibitem{Sundin:2008vt}
P.~Sundin, ``{The {$\ads_4 \times \C P^3$} string and its Bethe equations in
  the near plane wave limit},'' {\em JHEP} {\bf 02} (2009) 046,
\href{http://www.arXiv.org/abs/0811.2775}{{\tt 0811.2775}}.

\bibitem{Kristjansen:2008ib}
C.~Kristjansen, M.~Orselli, and K.~Zoubos, ``{Non-planar ABJM Theory and
  Integrability},'' {\em JHEP} {\bf 03} (2009) 037,
\href{http://www.arXiv.org/abs/0811.2150}{{\tt 0811.2150}}.

\bibitem{Zwiebel:2009vb}
B.~I. Zwiebel, ``{Two-loop Integrability of Planar N=6 Superconformal Chern-
  Simons Theory},'' {\em J. Phys.} {\bf A42} (2009) 495402,
\href{http://www.arXiv.org/abs/0901.0411}{{\tt 0901.0411}}.

\bibitem{Minahan:2009te}
J.~A. Minahan, W.~Schulgin, and K.~Zarembo, ``{Two loop integrability for
  Chern-Simons theories with N=6 supersymmetry},'' {\em JHEP} {\bf 03} (2009)
  057,
\href{http://www.arXiv.org/abs/0901.1142}{{\tt 0901.1142}}.

\bibitem{Ahn:2008aa}
C.~Ahn and R.~I. Nepomechie, ``{N=6 super Chern-Simons theory S-matrix and
  all-loop Bethe ansatz equations},'' {\em JHEP} {\bf 09} (2008) 010,
\href{http://www.arXiv.org/abs/0807.1924}{{\tt 0807.1924}}.

\bibitem{Gromov:2009tv}
N.~Gromov, V.~Kazakov, and P.~Vieira, ``{Integrability for the Full Spectrum of
  Planar AdS/CFT},''
\href{http://www.arXiv.org/abs/0901.3753}{{\tt 0901.3753}}.

\bibitem{Callan:2003xr}
J.~Callan, Curtis~G. {\em et al.}, ``Quantizing string theory in {$\ads_5
  \times S^5$}: Beyond the pp- wave,'' {\em Nucl. Phys.} {\bf B673} (2003)
  3--40,
\href{http://www.arXiv.org/abs/hep-th/0307032}{{\tt hep-th/0307032}}.

\bibitem{Callan:2004uv}
J.~Callan, Curtis~G., T.~McLoughlin, and I.~Swanson, ``{Holography beyond the
  Penrose limit},'' {\em Nucl. Phys.} {\bf B694} (2004) 115--169,
\href{http://www.arXiv.org/abs/hep-th/0404007}{{\tt hep-th/0404007}}.

\bibitem{Berenstein:2002jq}
D.~Berenstein, J.~M. Maldacena, and H.~Nastase, ``Strings in flat space and pp
  waves from {$\CN = 4$} super {Yang Mills},'' {\em JHEP} {\bf 04} (2002) 013,
\href{http://www.arXiv.org/abs/hep-th/0202021}{{\tt hep-th/0202021}}.

\bibitem{Arutyunov:2004vx}
G.~Arutyunov, S.~Frolov, and M.~Staudacher, ``Bethe ansatz for quantum
  strings,'' {\em JHEP} {\bf 10} (2004) 016,
\href{http://www.arXiv.org/abs/hep-th/0406256}{{\tt hep-th/0406256}}.

\bibitem{Gomis:2008jt}
J.~Gomis, D.~Sorokin, and L.~Wulff, ``{The complete {$\ads_4 \times \C P^3$}
  superspace for the type IIA superstring and D-branes},'' {\em JHEP} {\bf 03}
  (2009) 015,
\href{http://www.arXiv.org/abs/0811.1566}{{\tt 0811.1566}}.

\bibitem{Grassi:2009yj}
P.~A. Grassi, D.~Sorokin, and L.~Wulff, ``{Simplifying superstring and D-brane
  action in the {$\ads_4 \times \C P^3$} superbackground},'' {\em JHEP} {\bf
  08} (2009) 060,
\href{http://www.arXiv.org/abs/0903.5407}{{\tt 0903.5407}}.

\bibitem{Cvetic:1999zs}
M.~Cvetic, H.~Lu, C.~N. Pope, and K.~S. Stelle, ``{T-Duality in the
  Green-Schwarz Formalism, and the Massless/Massive IIA Duality Map},'' {\em
  Nucl. Phys.} {\bf B573} (2000) 149--176,
\href{http://www.arXiv.org/abs/hep-th/9907202}{{\tt hep-th/9907202}}.

\bibitem{futurework}
D.~Astolfi, V.~G.~M. Puletti, G.~Grignani, T.~Harmark, and M.~Orselli, ``Work
  in progress''.

\bibitem{Nilsson:1984bj}
B.~E.~W. Nilsson and C.~N. Pope, ``{Hopf fibration of eleven-dimensional supergravity},'' {\em Class. Quant. Grav.} {\bf 1} (1984)
499.

\bibitem{Sugiyama:2002tf}
K.~Sugiyama and K.~Yoshida, ``{Type IIA string and matrix string on pp-wave},''
  {\em Nucl. Phys.} {\bf B644} (2002) 128--150,
\href{http://www.arXiv.org/abs/hep-th/0208029}{{\tt hep-th/0208029}}.

\bibitem{Hyun:2002wu}
S.-j. Hyun and H.-j. Shin, ``{N = (4,4) type IIA string theory on pp-wave
  background},'' {\em JHEP} {\bf 10} (2002) 070,
\href{http://www.arXiv.org/abs/hep-th/0208074}{{\tt hep-th/0208074}}.

\bibitem{Nishioka:2008gz}
T.~Nishioka and T.~Takayanagi, ``{On Type IIA Penrose Limit and N=6
  Chern-Simons Theories},''
\href{http://www.arXiv.org/abs/0806.3391}{{\tt 0806.3391}}.

\bibitem{McLoughlin:2008he}
T.~McLoughlin, R.~Roiban, and A.~A. Tseytlin, ``{Quantum spinning strings in
  {$\ads_4 \times \C P^3$}: testing the Bethe Ansatz proposal},'' {\em JHEP}
  {\bf 11} (2008) 069,
\href{http://www.arXiv.org/abs/0809.4038}{{\tt 0809.4038}}.

\bibitem{Sundin:2009zu}
P.~Sundin, ``{On the worldsheet theory of the type IIA {$\ads_4 \times \C P^3$}
  superstring},''
\href{http://www.arXiv.org/abs/0909.0697}{{\tt 0909.0697}}.

\bibitem{Bena:2003wd}
I.~Bena, J.~Polchinski, and R.~Roiban, ``{Hidden symmetries of the {$\ads_5
  \times S^5$}superstring},'' {\em Phys. Rev.} {\bf D69} (2004) 046002,
\href{http://www.arXiv.org/abs/hep-th/0305116}{{\tt hep-th/0305116}}.

\bibitem{Fre:2008qc}
P.~Fre and P.~A. Grassi, ``{Pure Spinor Formalism for {Osp}(N|4)
  backgrounds},''
\href{http://www.arXiv.org/abs/0807.0044}{{\tt 0807.0044}}.

\bibitem{Rashkov:2008rm}
R.~C. Rashkov, ``{A note on the reduction of the {$\ads_4 \times \C P^3$}
  string sigma model},'' {\em Phys. Rev.} {\bf D78} (2008) 106012,
\href{http://www.arXiv.org/abs/0808.3057}{{\tt 0808.3057}}.

\bibitem{Dukalski:2009pr}
M.~Dukalski and S.~J. van Tongeren, ``{On fermionic reductions of the {$\ads_4
  \times \C P^3$} superstring},'' {\em Phys. Rev.} {\bf D80} (2009) 046005,
\href{http://www.arXiv.org/abs/0906.0706}{{\tt 0906.0706}}.

\bibitem{Bak:2009mq}
D.~Bak, H.~Min, and S.-J. Rey, ``{Generalized Dynamical Spin Chain and 4-Loop
  Integrability in N=6 Superconformal Chern-Simons Theory},''
\href{http://www.arXiv.org/abs/0904.4677}{{\tt 0904.4677}}.
%
J.~A. Minahan, O.~Ohlsson~Sax, and C.~Sieg, ``{Magnon dispersion to four loops
  in the ABJM and ABJ models},''
\href{http://www.arXiv.org/abs/0908.2463}{{\tt 0908.2463}}.
%
D.~Bak, H.~Min, and S.-J. Rey, ``{Integrability of N=6 Chern-Simons Theory at
  Six Loops and Beyond},''
\href{http://www.arXiv.org/abs/0911.0689}{{\tt 0911.0689}}.
%
G.~Papathanasiou and M.~Spradlin, ``{Two-Loop Spectroscopy of Short ABJM
  Operators},''
\href{http://www.arXiv.org/abs/0911.2220}{{\tt 0911.2220}}.
%
G.~Grignani, T.~Harmark, M.~Orselli, and G.~W. Semenoff, ``{Finite size Giant
  Magnons in the string dual of N=6 superconformal Chern-Simons theory},'' {\em
  JHEP} {\bf 12} (2008) 008,
\href{http://www.arXiv.org/abs/0807.0205}{{\tt 0807.0205}}.
%
D.~Bombardelli and D.~Fioravanti, ``{Finite-Size Corrections of the
  {$\mathbb{CP}^3$} Giant Magnons: the L\'{u}scher terms},'' {\em JHEP} {\bf
  07} (2009) 034,
\href{http://www.arXiv.org/abs/0810.0704}{{\tt 0810.0704}}.
%
T.~Lukowski and O.~O. Sax, ``{Finite size giant magnons in the SU(2) x SU(2)
  sector of {$\ads_4 \times \C P^3$}},'' {\em JHEP} {\bf 12} (2008) 073,
\href{http://www.arXiv.org/abs/0810.1246}{{\tt 0810.1246}}.
%
D.~Berenstein and D.~Trancanelli,
  ``Three-dimensional N=6 SCFT's and their membrane dynamics,''
  Phys.\ Rev.\  D {\bf 78}, 106009 (2008), \href{http://www.arXiv.org/abs/0808.2503}{{\tt 0808.2503}} 
%
  D.~Trancanelli,
  ``Emergent geometry in N=6 Chern-Simons-matter theory,'', \href{http://www.arXiv.org/abs/0904.0449}{{\tt 0904.0449}}
%
C.~Ahn and P.~Bozhilov, ``{Finite-size Effect of the Dyonic Giant Magnons in
  N=6 super Chern-Simons Theory},'' {\em Phys. Rev.} {\bf D79} (2009) 046008,
\href{http://www.arXiv.org/abs/0810.2079}{{\tt 0810.2079}}.
%
M.~C. Abbott and I.~Aniceto, ``{Giant Magnons in {$\ads_4 \times \C P^3$}:
  Embeddings, Charges and a Hamiltonian},''
\href{http://www.arXiv.org/abs/0811.2423}{{\tt 0811.2423}}.
%
C.~Kalousios, M.~Spradlin, and A.~Volovich, ``{Dyonic Giant Magnons on
  {$CP^3$}},'' {\em JHEP} {\bf 07} (2009) 006,
\href{http://www.arXiv.org/abs/0902.3179}{{\tt 0902.3179}}.
%
C.~Kalousios, C.~Vergu, and A.~Volovich, ``{Factorized Tree-level Scattering in
  {$\ads_4 \times \C P^3$}},'' {\em JHEP} {\bf 09} (2009) 049,
\href{http://www.arXiv.org/abs/0905.4702}{{\tt 0905.4702}}.
%
M.~C. Abbott, I.~Aniceto, and O.~O. Sax, ``{Dyonic Giant Magnons in $CP^3$:
  Strings and Curves at Finite J},'' {\em Phys. Rev.} {\bf D80} (2009) 026005,
\href{http://www.arXiv.org/abs/0903.3365}{{\tt 0903.3365}}.
%
Y.~Hatsuda and H.~Tanaka, ``{Scattering of Giant Magnons in $CP^3$},''
\href{http://www.arXiv.org/abs/0910.5315}{{\tt 0910.5315}}.

\bibitem{McLoughlin:2008ms}
T.~McLoughlin and R.~Roiban, ``{Spinning strings at one-loop in {$\ads_4 \times
  \C P^3$}},'' {\em JHEP} {\bf 12} (2008) 101,
\href{http://www.arXiv.org/abs/0807.3965}{{\tt 0807.3965}}.

\bibitem{Alday:2008ut}
L.~F. Alday, G.~Arutyunov, and D.~Bykov, ``{Semiclassical Quantization of
  Spinning Strings in {$\ads_4 \times \C P^3$}},'' {\em JHEP} {\bf 11} (2008)
  089,
\href{http://www.arXiv.org/abs/0807.4400}{{\tt 0807.4400}}.

\bibitem{Krishnan:2008zs}
C.~Krishnan, ``{{$\ads_4$}/CFT$_3$ at One Loop},'' {\em JHEP} {\bf 09} (2008)
  092,
\href{http://www.arXiv.org/abs/0807.4561}{{\tt 0807.4561}}.

\bibitem{Bandres:2009kw}
M.~A. Bandres and A.~E. Lipstein, ``{One-Loop Corrections to Type IIA String
  Theory in {$\ads_4 \times \C P^3$}},''
\href{http://www.arXiv.org/abs/0911.4061}{{\tt 0911.4061}}.

\bibitem{Zarembo:2009au}
K.~Zarembo, ``{Worldsheet spectrum in AdS(4)/CFT(3) correspondence},''
\href{http://www.arXiv.org/abs/0903.1747}{{\tt 0903.1747}}.

\bibitem{Bergman:2009zh}
O.~Bergman and S.~Hirano, ``{Anomalous radius shift in {$\ads_4 \times \C
  P^3$}},'' {\em JHEP} {\bf 07} (2009) 016,
\href{http://www.arXiv.org/abs/0902.1743}{{\tt 0902.1743}}.

\bibitem{Gromov:2008bz}
N.~Gromov and P.~Vieira, ``{The AdS4/CFT3 algebraic curve},'' {\em JHEP} {\bf
  02} (2009) 040,
\href{http://www.arXiv.org/abs/0807.0437}{{\tt 0807.0437}}.

\bibitem{Gromov:2008fy}
N.~Gromov and V.~Mikhaylov, ``{Comment on the Scaling Function in {$\ads_4
  \times \C P^3$}},'' {\em JHEP} {\bf 04} (2009) 083,
\href{http://www.arXiv.org/abs/0807.4897}{{\tt 0807.4897}}.

\bibitem{Metsaev:1998it}
R.~R. Metsaev and A.~A. Tseytlin, ``{Type {IIB} superstring action in {$\ads_5
  \times S^5$} background},'' {\em Nucl. Phys.} {\bf B533} (1998) 109--126,
\href{http://www.arXiv.org/abs/hep-th/9805028}{{\tt hep-th/9805028}}.

\bibitem{Kallosh:1998zx}
R.~Kallosh, J.~Rahmfeld, and A.~Rajaraman, ``{Near horizon superspace},'' {\em
  JHEP} {\bf 09} (1998) 002,
\href{http://www.arXiv.org/abs/hep-th/9805217}{{\tt hep-th/9805217}}.

\bibitem{Cagnazzo:2009zh}
A.~Cagnazzo, D.~Sorokin, and L.~Wulff, ``{String instanton in AdS(4)xCP(3)},''
\href{http://www.arXiv.org/abs/0911.5228}{{\tt 0911.5228}}.

\bibitem{Uvarov:2009nk}
D.~V.~Uvarov,
``Light-cone gauge Hamiltonian for $AdS_4$ x $CP^3$ superstring,'', \href{http://www.arXiv.org/abs/0912.1044}{{\tt 0912.1044}}

\bibitem{Green:1983wt}
M.~B. Green and J.~H. Schwarz, ``{Covariant Description of Superstrings},''
  {\em Phys. Lett.} {\bf B136} (1984)
367--370.

\bibitem{Green:1983sg}
M.~B. Green and J.~H. Schwarz, ``{Properties of the Covariant Formulation of
  Superstring Theories},'' {\em Nucl. Phys.} {\bf B243} (1984)
285.

\bibitem{Grisaru:1985fv}
M.~T. Grisaru, P.~S. Howe, L.~Mezincescu, B.~Nilsson, and P.~K. Townsend,
  ``{N=2 Superstrings in a Supergravity Background},'' {\em Phys. Lett.} {\bf
  B162} (1985)
116.

\bibitem{Bertolini:2002nr}
M.~Bertolini, J.~de~Boer, T.~Harmark, E.~Imeroni, and N.~A. Obers, ``Gauge
  theory description of compactified pp-waves,'' {\em JHEP} {\bf 01} (2003)
  016,
\href{http://www.arXiv.org/abs/hep-th/0209201}{{\tt hep-th/0209201}}.

\bibitem{Harmark:2006ta}
T.~Harmark and M.~Orselli, ``Matching the {Hagedorn} temperature in
  {AdS/CFT},'' {\em Phys. Rev.} {\bf D74} (2006) 126009,
\href{http://www.arXiv.org/abs/hep-th/0608115}{{\tt hep-th/0608115}}.

\bibitem{Harmark:2008gm}
T.~Harmark, K.~R. Kristjansson, and M.~Orselli, ``{Matching gauge theory and
  string theory in a decoupling limit of AdS/CFT},''
\href{http://www.arXiv.org/abs/0806.3370}{{\tt 0806.3370}}.

\bibitem{Astolfi:2008yw}
D.~Astolfi, G.~Grignani, T.~Harmark, and M.~Orselli, ``{Finite-size corrections
  to the rotating string and the winding state},'' {\em JHEP} {\bf 08} (2008)
  099,
\href{http://www.arXiv.org/abs/0804.3301}{{\tt 0804.3301}}.

\end{thebibliography}

\addcontentsline{toc}{section}{References}

\providecommand{\href}[2]{#2}\begingroup\raggedright\endgroup

\end{document}